\documentclass[twocolumn]{aastex631}
\usepackage{amsmath}
\usepackage{wasysym}
\usepackage{hyperref}
\usepackage{xcolor}
\usepackage{url}
\usepackage[version=4]{mhchem}

\def\gtaprx{ \mathrel{ \vcenter{
      \offinterlineskip \hbox{$>$}
      \kern 0.3ex \hbox{$\sim$}    } } }

\def\ltaprx{ \mathrel{ \vcenter{
      \offinterlineskip \hbox{$<$}
      \kern 0.3ex \hbox{$\sim$}    } } }
      
\def\aj{{AJ}}
\def\apj{{ApJ}}
\def\apjs{{ApJS}}
\def\apjl{{ApJL}}

\def\aap{{A\&A}}
\def\pasp{{PASP}}

\shorttitle{Dynamic Small Database Design for SpExoDisks }
\shortauthors{Wheeler, Hinkel, \& Banzatti}

\begin{document}

\title{Database Design for SpExoDisks: A Database \& Web Portal for Spectra of Exoplanet-Forming Disks}

\author[0000-0001-5563-6987]{Caleb H. Wheeler III}
\affiliation{LIGO Livingston Observatory, Livingston, LA 70754, USA}

\author[0000-0003-0595-5132]{Natalie R.\ Hinkel}
\affiliation{Louisiana State University, Department of Physics and Astronomy, 202 Nicholson Hall, Baton Rouge, LA 70803, USA}

\author[0000-0003-4335-0900]{Andrea Banzatti}
\affiliation{Department of Physics, Texas State University, 749 N Comanche Street, San Marcos, TX 78666, USA}

\correspondingauthor{Caleb H. Wheeler III}
\email{chw3k5@gmail.com}

\begin{abstract}
Data access -- or the availability of new and archival data for use by the larger community -- is key for scientific advancement. How data is presented, searched, and formatted determines accessibility and it can be difficult to find a solution that fits the needs of a given subdiscipline. We present a generalized roadmap for developing a specialty astronomy database with web application based on the development of the SpExoDisks (Spectra of Exoplanet forming Disks) database (\url{spexodisks.com}), which provides infrared spectra of protoplanetary disks. Expertise in an astronomy subdiscipline can provide two necessary components for creating a database: access to a large volume of specialized data and knowledge of how that data should be presented to the community. However, there are a variety of steps and decisions for database development that can fall outside astronomy expertise. Here we offer generalized discussions on design and process that are accompanied by real-world examples from the SpExoDisks developer team and website. Starting from the database portal design and data organization, we demonstrate on-demand data distribution and query using publicly accessible database software. These systems support interactive visualizations such that users can explore spectra directly from their browsers. We also offer details that show how the technical concepts in SpExoDisks are implemented, particularly emphasizing sustainability and long-term management of the codebase and processes. Finally, we illustrate the utility that a specialty website can offer to the community by providing a specific example of how the combined spectra from SpExoDisks can enhance our understanding of protoplanetary disks.

\end{abstract}


\section{Introduction}
We present the {\textbf{Sp}ectra for \textbf{Exo}planet-Forming \textbf{Disks}, or SpExoDisks, database portal, which hosts spectra and purpose-built tools to serve the needs of a specific science community interested in infrared spectra of protoplanetary disks, especially those with exoplanets (\url{spexodisks.com}). The SpExoDisks database\footnote{To disambiguate the uses of the word ``database", a qualifier, such as portal, software, or table will always accompany it.} portal enables users to view and sort data in an online browser application as well as download specific data. The codebase that comprises SpExoDisks is designed to grow and change in lockstep with the science it supports. 

In general, the amount and complexity of science data continues to increase. Libraries of existing data can become less accessible or inaccessible over time as technologies become incompatible. Fortunately, there are a few large, excellent science database portals (websites). For example, SIMBAD \citep{Wenger2000}, Gaia \citep{Gaia2016b, Gaia2023j}, TIC \citep{StassunOelkers_2019}, and HITRAN \citep{hitran20} provide data and help drive discovery through big science. In addition, these large database portals create an ecosystem that allows SpExoDisks to be updated dynamically: as we receive new spectra from scientific observations, we query other databases portals to associate new data with existing data and provide a context for scientific qualification. Small database portals like SpExoDisks enable astronomers to sort through the increasing deluge of available data and provide a place for the future generations of scientists to learn. 

The processes used by SpExoDisks and discussed in the present work are a direct evolution of the Hypatia Catalog (\url{hypatiacatalog.com}), the largest astronomy database of elemental abundances within stars near to the Sun \citep[or within 500pc,][]{Hinkel14}. The two databases have been developed in an alternating series with one another: beginning with Hypatia, then expanding and improving that framework to create SpExoDisks, which fed back into new innovations for an updated Hypatia Catalog. Like SpExoDisks, the Hypatia Catalog is a database that unites different sources to serve a specific community. Both are designed to respond and change with audiences and communities, even though the audiences for stellar chemical trends and infrared spectra for stellar disks may be different. 

The present work discusses the design and intent of the SpExoDisks online data portal. While the examples are specific to the SpExoDisks project, we present from the perspective of experience and determining what is useful for building database portals that solve the specific needs of a given scientific community. SpExoDisks does not try to solve all astronomy database problems; instead, it focuses on delivering spectra to be searched and viewed before being downloaded with data-extraction tools and documentation. This limited scope allows SpExoDisks to be maintained and upgraded for and with the community it serves. Database portals scoped for a specific community can be a sustainable tool to reduce complexity and provide data access to wider communities while contributing to an astronomy data ecosystem.  

In this paper, we explain the logic and decisions that were necessary to consider when building a niche astronomy database portal, SpExoDisks, using the experience and expertise of the developer team to outline best practices and caveats. In \S \ref{s.begin-portal}, we consider the goals of the database portal including the primary data and secondary contextual data, in addition to how that data must be organized, standardized, and verified. In \S \ref{s.database_software}, we go over database software in terms of query efficiency for the developers and users, as well as API access. In \S \ref{s.data_interactivity_accessibility}, we consider how to visualize the primary and secondary data while taking into account accessibility. We also focus on how to provide data either for direct download or through programmatic code/scripts. In \S \ref{s.technical}, we give details as to the software that was used by the SpExoDisks team to securely develop the infrastructure on a variety of computer architectures, as well as the specific setup for the data science processing pipeline and public server configuration (see the open-source code at \url{https://github.com/spexod/Portal}). Finally, in \S \ref{s.science_application}, we provide a science application that shows the utility and breadth of the SpExoDisks database portal and the ways that it will help strengthen and evolve the protoplanetary disk community.

\section{Beginning a Database Portal}\label{s.begin-portal}

Designing a database portal of any type requires making hundreds of design choices. Designing or building processes that \textit{may be} useful in the future or are too abstract can drastically limit progress. For this reason, we recommend starting by defining your project's goals (\S \ref{s.goals}). These goals will help determine the project's scope, establishing which tasks are relevant to the project while also considering what the project can accomplish in one, three, or ten years. Core goals, or the minimum requirements for success, can be prioritized over reach goals, such as ancillary features and/or long-term upgrades. 

Next, the specific data needs to be designed to serve the project's defined goals. The data will include the primary, core information within the database portal, e.g. IR spectra for SpExoDisks or elemental abundances for Hypatia Catalog, but it may also include secondary data that provides context and supports the exploration of the primary data (\S \ref{s.data_context}). For a given database software (i.e. specialized software for data handling and query), it's important to design tables or other data-viewing objects that support the project goals regarding values, references, and other data types (\S \ref{s.data_organization}). This can also include the data relationships across tables with the database software. 

Once the project's data structure is organized, the next step is to consider what processes and calculations will be needed to go from the raw input data's format to the structure required to support the project (\S \ref{s.standardization_and_verification}). We refer to it as a data processing pipeline and this stage can be a substantial fraction of the project's effort. However, if we can track pipeline processes to core project goals, judicious trades can be made regarding spending time and other resources for developing new pipeline processes. 

\subsection{Database Portal Goals}\label{s.goals}

A common goal when developing a database portal is to create one that is easy to understand, fast, and available. However, accomplishing all three can be difficult because scientific data is often incomplete and irregular. For example, a star may have many or no values available for its effective temperature. Scientific database portals must also do more than report mean values; for each value, they must include references to literature, observational source materials, and they additionally require reported units and errors.

How data will be accessed (or used) informs the project's data structure. That is why this is the first question that the SpExoDisks team had to address before developing a database portal. Our main goal is to deliver per-observation spectra that could be viewed and downloaded while also allowing the spectra to be navigated using per-object data like multiple star names or measured values. Our project's primary data type, observed spectra, is supported by context from other secondary data types. 

Let us consider the way that the SpExoDisks project supports the main goal. Our portal's primary data product is spectra, each stored in the database software as a table with wavelength, flux, and flux error columns. Visual inspection of the spectra from an expert is fundamental to ensure the data is scientifically valid. Therefore, the processing software that stores the spectra also includes a plotting tool with molecular lines and other significant context for association and debugging.

An example of how we support our secondary goals can be seen in how we give our users the ability to explore trends. This forward-looking process considers the database portal as a whole and not a collection of individual spectra. Some metadata is per spectra, like science analysis that provides values of flux calculated for specific spectral lines, the quantum states of molecular lines, or simply the time of observation. Some data is per star (each star can have one or many spectra in SpExoDisks), such as the star's distance, effective temperature, or its right ascension and declination coordinates. The secondary data is crucial to providing additional resources that can help the users answer key science questions quickly and efficiently.

\subsection{Automated Data Contextualization}\label{s.data_context}

Hosting a collection of observations is relatively simple; in about a day, a website could be launched on a cloud service to serve data files and their directories. However, a database portal can be more than a collection of individual observations. It can also provide the necessary and diverse contexts that allow users to search and retrieve data and, perhaps more importantly, see what is missing from an existing dataset. The context requirements for data depend on the type of questions users will choose; what questions does the database portal hope to answer? Automatically creating data associations and acquiring additional context within the data processing pipeline is paramount for the long-term viability and continuity of a database portal. 

In astronomy, we are fortunate to have many large database portals that can help us give context to other astronomy data. Three database portals are specifically used for contextualizing spectra for the SpExoDisks project: SIMBAD, Gaia, the Tess Input Catalog (TIC). These database portals provide an Application Program Interface (API) that allows users to automate data query and retrieval. In particular, the SIMBAD database’s API is a primary tool used to link two observations of the same object that are identified by different names (see the star name problem discussed in Appendix \ref{s.star_name_problem}). The SIMBAD, Gaia, and TIC database portals also provide information like right ascension/declination, distance, and effective stellar temperature.

Automatic data contextualization is a major strength of SpExoDisks; it allows us to take raw spectra and deliver them to our data portal in a fast process that lasts less than 24 hours for users to view and download immediately. The SpExoDisks data processing pipeline self-contextualizes using only the star names and parameter data provided in the original file from the observation. The automatic data contextualization process is a facet of our data curation; the data processing pipeline raises exceptions where the process exits unsuccessfully when minimum criteria are unmet. We found that raising errors that must be fixed forces our team to address those issues or remove the data from processing before it is uploaded to the public website. This is one of the ways our data meets minimum curation standards for uniformity (the curation process is discussed in \S \ref{s.standardization_and_verification}).

In addition, the SpExoDisks data processing pipeline saves data from other database portals locally, so the local data is first checked before querying external database portals. There are several reasons to take this approach, but most importantly, it is about not overusing the community's database portal resources. Namely, it is not necessary to use database portal resources to ask the same questions for all stars in SpExoDisks hundreds of times only to get the same answers. Another practical reason to save data locally, is speed; even the fastest databases, like Gaia, are still many times slower than simply ready a locally available file or an existing table database software. Deleting all existing stellar names and reference data in the data processing pipeline automatically triggers the pipeline to remake new reference files. However, this process can take several hours for SIMBAD.

However, the drawback to using local data is staleness. Astronomical knowledge can undergo rapid changes; stellar parameters are continually being updated as improved pipelines and new measurements are available. A named identifier given to what appeared to be a single star may later be updated after being reclassified as a multi-star system. To avoid staleness, the local data used by the SpExoDisks database portal must be deleted to trigger a fresh download of the latest information. For SpExoDisks, this is done once a year in January.

The tools that both SpExoDisks and Hypatia Catalog \citep{Hinkel14} use for data discovery and caching -- specifically from the SIMBAD\footnote{\url{https://simbad.u-strasbg.fr/simbad/sim-fid}}, Gaia\footnote{\url{https://gea.esac.esa.int/archive/}}, TIC\footnote{\url{https://mast.stsci.edu/portal/Mashup/Clients/Mast/Portal.html}}, and the NASA Exoplanet Archive\footnote{\url{https://exoplanetarchive.ipac.caltech.edu/}} -- are freely available for download and collaboration as the Python package \texttt{autostar}\footnote{\url{https://github.com/chw3k5/autostar}} or installed via pypi.org: \texttt{pip install autostar}. This repository lacks documentation and tools to make it user-friendly. However, it does integrate astropy \citep{Astropy22} queries with simple databases made of human-readable CSV and PSV files. With more users/collaborators and/or with grants from new databases, we are open to improving this code or adapting the idea to new applications.

\subsection{Data Organization}\label{s.data_organization}

The next step of database portal design is identifying how to sort and contextualize your main, primary data (as opposed to the secondary contextual data). Since spectra are the primary data in SpExoDisks, we could have started by simply sorting all the input file paths in alphabetical order and giving them a unique identifier index (a counter assigned based on the filename order). However, because of inconsistencies with the file names across all spectra, we wanted a spectrum ID that included the observed star's name and other identifying information so spectrum data could be understood without opening the file. The fundamental unique key of the SpExoDisks database portal is the spectrum ID, but it depends first on another crucial key, the star ID, a unique string that maps to a single stellar system.

Unfortunately, the same star can be known by many names across catalogs of stellar names, datasets, and surveys. Correctly linking names from major stellar catalogs for the SpExoDisks dataset was a critical data science problem since we have diverse input names formats from eight instruments so far. When the SpExoDisks data processing pipeline encounters a spectrum with an unknown stellar name, a data collection process is triggered to determine if this object has a known star name within SpExoDisks or if a new star name entry is required (more on this process in \S \ref{s.data_flow}). In the case of a new name entry, additional data retrieval is triggered to determine the available stellar names. Each new spectrum encountered by the data processing pipeline joins or creates a parent single-star data object. We also chose to make separate single-star objects for multi-star systems, compared to those designed for individual measurements of the constituent stars. For a more lengthy discussion of the star name problem, see Appendix \ref{s.star_name_problem}; for now, we suppose spectrum and all metadata can be associated with a unique name for a star or stellar system, called a star ID. Within SpExoDisks, we require each spectrum to have a star or stellar system name, a scientific reference, and an observation date (see top right of Fig. \ref{fig:db_tables}). Once all data is associated with a star or spectrum, pipeline tools perform further processing for all contextual data associated with a single star, all data for a single spectra, and the whole of all available data.

Each spectrum in SpExoDisks requires a unique identifier for the data processing pipeline. We chose to make an identifier that would be helpful to read and sort in a direct structure. We combined the observing instrument name, the wavelength range (in nanometers rounded to whole integers), and the star's name to create the spectrum ID, for example ``crires\_2338nm\_2396nm\_IRAS\_10418minus5931". However, this spectrum ID configuration does not guarantee a unique name or a name that is persistent with configuration changes in the processing pipeline. We resolved the issue by creating a spectrum ID registration system within the persistent MySQL\footnote{MySQL (\url{mysql.com}) is an open-source database management software that allows data to be filtered, organized, and accessed across tables.} online storage (see \S \ref{s.public-server}) where a given spectrum ID is registered to an input filename and is never allowed to be associated with any other file. A simpler process might have included the observation time in the spectrum handle or other information that enforced uniqueness without requiring registrations. One visible application of the spectrum ID is in the URL for a given spectrum in the SpExoDisks data portal, per the same example as before: \url{spexodisks.com/ExploreData/crires_2338nm_2396nm_IRAS_10418minus5931}. Note that these handles have star names that replace special characters to meet the standards for URL patterns, as discussed in the last paragraph of Appendix \ref{s.star_name_problem}.

\begin{figure*} 
    \centering
    \includegraphics[width=\linewidth]{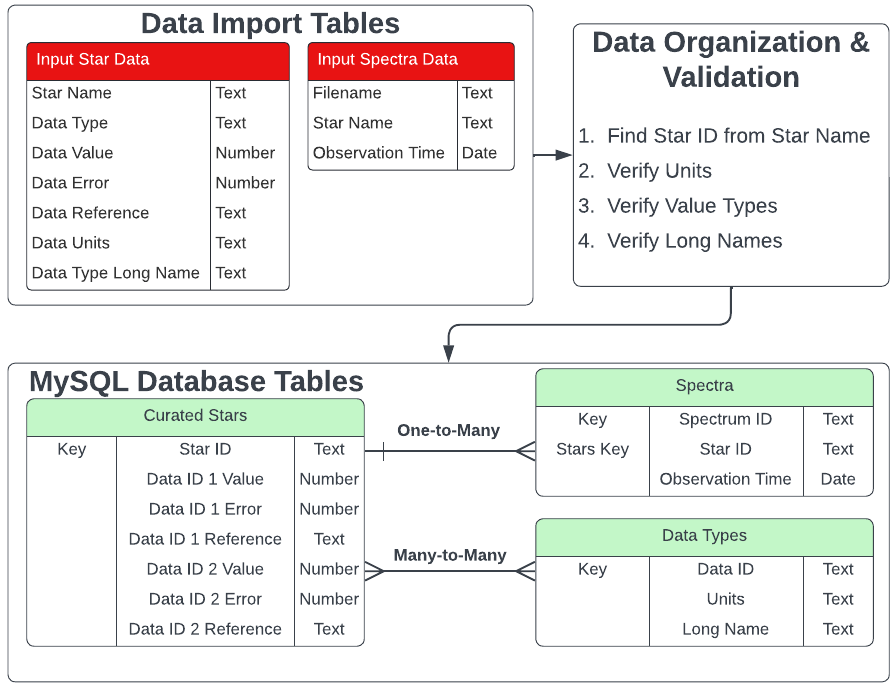}
    \caption{
    A schematic of the data organization and validation process for the SpExoDisks database portal. Before being imported, stellar and spectral data are stored in tables (shown with white text on a red background), where the data's column name and type are listed. Before read-in, tables have no enforced rules or known relationships between them. Data validation (top right, numbered steps) makes associations between data components by linking name data to standardized keys and enforcing rules important for scientific validity. After validation, new tables are created in the MySQL database software (shown with black text on a green background) with linking relationships via key values. The tables created in the MySQL database software have known rules enforced. This provides consistent ordering and naming within each table and creates links between tables with known one-to-many and many-to-many relationships, as illustrated with the crow's feet notation arrows. The notation of \textit{Key} indicates that the column will contain unique key values to identify and link a given data record. In the ``Spectra" table, the \textit{Stars Key} indicates that the \textit{Star ID} column contains non-unique values that link to records in the ``Curated Stars" table.  
    }\label{fig:db_tables}
\end{figure*} 

\subsection{Data Standardization and Verification}\label{s.standardization_and_verification}

Most of \S \ref{s.data_organization} focused on assembling data from many parts in preparation for data collation within the processing pipeline. With all the raw data products gathered and linked, we next need to standardize the data and export it to the MySQL database software for use by a larger audience (see Fig. \ref{fig:db_tables}). The database portal can be considered a subset of the available data in the data processing pipeline. In contrast, the data processing pipeline can contain more data types than what is subsequently viewable from the online database portal. 

A database portal can limit the data provided to the user so that most data types and fields are complete. Allowing many special data types (e.g., data provided in erg vs Joule or $\mu$m vs nm vs \AA) will increase code complexity, making the database portal less manageable since the effort needed to collect, compare, and update the constituent data components will only increase over time. These are very tempting pitfalls for scientists who want to do the best job possible by delivering all possible data. Unfortunately, there is no way to provide everything that all users will ever need. We recommend a consistent method of delivering the data that will allow users to make the necessary modifications for their science within their own processes.

A database portal can provide the main results quickly, including measured values, units, errors, and references (``Data Import Tables" in Fig. \ref{fig:db_tables}) that enable users to go directly to scientific papers to find additional results, special contexts, and caveats. In the SpExoDisks database portal, all data has custom data objects that contain this information (see Appendix \ref{s.custom_data_objects}). We have chosen to support different upper and lower error values for stellar parameters. To make the shape of the public database uniform, we convert all stellar parameters with single plus/minus error values to upper and lower values. 

In the SpExoDisks database portal, we enforce strict unit controls. Only one unit type is allowed for a given data type. Data types and units are specified in a configurable file that is also used to determine 1) if data values should be exported from the pipeline to the database software, 2) the short name of the data type when displayed on the website, and 3) the long name of the data that is shown within plots. Converting all data to have the same units makes the database portal searchable and comparable for parameter values (see \S \ref{s.query_efficiency_in_database_software}). For example, we can quickly sort data by any stellar parameter or report spectra in order of minimum wavelength values in a single step since all values for minimum wavelength are first converted to micrometers. 

While the dataset available to the data processing pipeline is abstractly shaped (like a tree with many different branches and leaves) in the same way as the dataset accessible from the MySQL database software, the pipeline does not contain repeated data values (only repeated references to a single value), while the database portal stores some repeated values. This is a purposeful choice from the paradigm where data storage is cheap while CPU and human-developer time are more expensive. For example, we do not ask the public database portal to do calculations to determine if the provided error represents upper and lower values by checking whether those values are reported separately. Instead, we have one column for upper error and one column for lower error. Stellar object data is stored in a table with one star per row with a unique star ID, and spectra metadata is stored in a table with one spectrum per row with a unique spectrum ID. Spectra and stellar data are stored separately so as not to make one very large table with stellar data repeated for any number of associated spectra (e.g., multiple spectra observations of the same star). To combine or join (a term borrowed from the MySQL database software) these tables, we associate the star ID for each spectrum with a row in the stellar parameters table, as illustrated by the crow's feet notation arrows in the ``MySQL Database Tables" in Fig. \ref{fig:db_tables}. While it will be discussed in more detail in \S \ref{s.technical}, it should be noted that other database software architectures do not require casting data into tables and instead allows a more flexible structure like MongoDB's document-oriented model.\footnote{\url{www.mongodb.com/}} In this model, documents are any number of string fields paired with values. Values include single types like strings, numbers, and dates, but additionally allow arrays of values and further levels for field-value objects. This allows the single document to directly contain the one-to-many relationships that might require multiple or repetitive tables in SQL tables. Like the SQL model of a table where each row is identifiable with a key, MongoDB's document-oriented model uses collections of documents where each has an identification key. However, the documents can store data in more flexible formats than the comparable SQL rows. 

Finally, the database portal should be conformal, meaning there is a single mapping for any incarnation of the raw data + data processing pipeline to the database portal (``MySQL Database Tables" in Fig. \ref{fig:db_tables}). As an example, consider the stellar parameter of distance to a star. In the data processing pipeline, we often collect many values for stellar distance (e.g., from Hipparcos, Gaia DR1/DR2/DR3, TIC, SIMBAD, etc.) for every star, creating an additional dimension of data to be curated. However, we simplify this wealth of data for the database portal and only report a single value to make the data easily searchable and streamline the user experience. To make the upload of data values conformal for stellar distance, we select the reported values using a ranked list of references in order of preference for a given data type. For example, we prefer Gaia DR3 to Gaia DR2 distance values. This is a critical curation step that should be performed by the principal scientist on a given project. 

Another important step to making a database portal conformal is to ensure the tables are presented in an intentional and reproducible order. For example, the spectrograph metadata table within the SpExoDisks database portal provides summary information about each spectrograph with at least one observed spectrum in the database portal. The order of the entries within this table sets the order of spectrograph summary data on the SpExoDisks homepage as well as the order in which instrument data columns are displayed on the ``Explore Data" page. Without conformal ordering, these data components would move around each time data was updated or require additional sorting and/or control code in the web portal's codebase. From the beginning of the SpExoDisks project, it was determined that all data processes and ordering should be controlled from the data processing pipeline to reduce the scope of knowledge needed to change the data.

Creating a database portal requires many decisions: Where will the data come from? How will it be maintained and updated? How will multidimensional data be reduced and a single value displayed? Because all of these decisions must be in support of the project's goals, it is important that they be supported by scientific justification to ensure consistency within the database portal as it continues to grow and develop. While this may seem obvious for some decisions (e.g., to prefer Gaia DR3 stellar distances to Gaia DR2 because they are the most up-to-date), it may not be the case for all (e.g., which star names to display as default). It is also imperative that these decisions be made evident to the user so that they understand the origin of the data they are using and can make informed choices regarding their science application.

\section{Database Software}\label{s.database_software}

The previous section discussed the data pipeline processes that deliver data and analysis in alignment with the goals of the database portal. We will continue the discussion as to what can be expected from the existing ecosystem of database software tools and techniques. Our aim is to host our finished database software in a system that is easy to understand, fast, and available online. While flexible Python database software may be ideal for research in a scientific group, we must also consider how this data will be accessed on a website. For a deployed database portal, we want something that can run on a database server (not only on personal computers), has sufficient access controls, and has been designed for remote deployment. In this case, a team might consider many popular database software like MongoDB, MySQL, PostgreSQL\footnote{\url{www.postgresql.org/}}, Oracle Live SQL\footnote{\url{www.oracle.com/database/technologies/oracle-live-sql/}}, or SQLite\footnote{\url{www.sqlite.org/}}. 

For the SpExoDisks project, we chose MySQL: a popular database software with a large community, a reliable history, great documentation, integration with Python, and a GUI application useful for navigating data to and issuing rare commands. We have been pleased with the way that MySQL hosts our primary spectrum data as well as the metadata that gives it context. Since we always return the full spectrum data, never a subsection, we could do well with a system that treats the spectrum as a file that is unpacked when received by the web portal codebase. However, if instead we were interested in retrieving a slice of the spectrum in wavelength, MySQL has excellent tooling for doing this efficiently. On the other hand, it seems there could be a better choice for hosting spectra from the perspective of upload time. Each spectrum is represented as a single table with three columns: wavelength, flux, and flux error. However, having the spectra accessible in a database software like MySQL allows us to do computationally efficient operations across many columns (i.e., of flux). Computationally intensive machine learning or modeling applications could be optimized for column-wise calculations directly in the database software, but triggered from a request from a Python API (see \ref{s.api_access_to_database}). 

\subsection{Query Efficiency in Database Software}\label{s.query_efficiency_in_database_software}

The organization of data within a given database software must consider how the data will be accessed most of the time. If a dataset is small and infrequently accessed, loading data directly from a CSV file in the final product may not significantly affect speed or user experience (users only see a presentation of the data, not the retrieval process). However, it is often worth considering how users will access data and then reduce the computations needed to get their requested data. Large, infrequently accessed database portals might decide to optimize for compactness. Many database portals will need substantial filtering to get users their requested information; such database portals can consider an optimization that includes indexing data values for faster comparison and data return. In addition, database portals may be supported by teams that expect to view and/or compose data in a specific format. In this case, the most sustainable database software for a project may be one optimized for understanding and readability. 

Early in the development of the SpExoDisks database portal, we used a relational database design to put the data in the third normal form (3NF). In 3NF, each table in the MySQL database software can uniquely identify each row with a key such that all data values for that row depend only on the key. This technique defines tree/branch relations when one table's keys link into another table's column names/values. For example, consider a database table that stores stellar parameter information. Suppose the first column is a unique key, \textit{star\_id}, while the second and third column stores the values of specific parameter types (for example, effective temperature and distance, respectively) denoted by the column name, \textit{param\_type1} and \textit{param\_type2}, respectively. A less compact table might have additional columns for the units of the parameters, e.g., \textit{units\_param\_type1} and \textit{units\_param\_type2}, where each value in the column is the same for all rows (i.e. listing `K' in reference to all values of stellar effective temperature and `pc' to all values of distance). The 3NF version of this data would be split into two tables: one with columns of \textit{star\_id}, \textit{param\_type1}, and \textit{param\_type2}; the other table with columns of \textit{param\_type} and \textit{units} where each parameter (e.g., effective temperature and distance) in a row is a key linking to a single value in the units column (such as `K' and `pc'), as illustrated by the crow's feet notation arrows in the ``MySQL Database Tables" in Fig. \ref{fig:db_tables}. In this example, the second table only has two rows, but more rows can be added for each additional parameter column that is included in the table with \textit{star\_id} as the key. In this way, no information is repeated, but there is enough information to link the tables together and recreate the original bulky, combined table.

The SQL syntax has been constructed such that combining data tables always occurs efficiently. On the other hand, Python database software, such as \texttt{pandas}\footnote{\url{pandas.pydata.org/}}, takes a different approach by providing more flexibility in ways to access data. For data operations, developers must learn to select between tools for inefficient queries (loops of many single queries to make an array) and more effective tools for array-based operations. Consequentially, the SpExoDisks team found that learning the SQL approach helped clarify how to write efficient \texttt{pandas} data pipeline processes. 

The SpExoDisks database portal contains 3NF data with per star data and per spectrum data separated into different tables to be joined later, as discussed in \S \ref{s.standardization_and_verification}. SQL tables in the 3NF form are compact because they only store data values in a single place, which can be especially important for storing large datasets. We compact SpExoDisks data to the 3NF at points in the data processing pipeline even when we plan to provide the data in a less compact form. This is because placing data in the 3NF makes it simple to check for consistency, minor typos, and other errors in star names and units (\S \ref{s.standardization_and_verification}).   

However, 3NF tables are not optimal from the perspective of computation efficiency. Doing many repeated (possibly nested) table-combining operations to construct a desired format can take a long time and confuse database portal users and developers. When database portals are not limited by data storage size, developers should consider a data architecture that requires less computation effort and will therefore return results faster. When fulfilling a user query, developers should also consider if filtering or reducing the returned values can be performed before table-joining computations to minimize the computations required. The SpExoDisks project uses a mixed approach for query efficiency and human readability. Some information that is periodically repeated, like a reference for a value's origin (e.g., a reference to Gaia DR3 for a stellar distance value) is allowed to be repeated in a per star table. However, some data in individual tables -- for example, separate tables tracking per parameter, per instrument, and whole database statistics -- significantly reduces repeated values in the more considerable per star and per spectrum tables. As a result, the database portal's API can directly distribute all the data on the website since the data has already been shaped for easy ingestion by those website components, such as the navigation table and spectrum plot in SpExoDisks. 

The SpExoDisks database software only uses a single way of arranging data values, what we call a single data structure. However, some use cases may call for multiple data structures, with each structure being an independent representation of all the available data. This can be useful for minimizing filtering computations to provide faster access to specific views or cuts of the data. We can imagine a hypothetical use of the SpExoDisks database portal as an input for modeling the dusty disks of stars, grouped by disk inclination angle. With the existing SpExoDisks data portal, we can return the disk inclination grouping using filtering computations for face-on, edge-on, and other inclination angles. In this case, we could also pre-filter the data into three similar data structures during the data science pipeline where filtering by disk inclination is done by pointing the API to the correct tables, rather than comparing many values in the database software. Any number of data structures can be uploaded from the same data processing pipeline.

\subsection{API Access to Database Software}\label{s.api_access_to_database}

The primary function of the SpExoDisks database portal's API is to provide direct, on-demand access to SpExoDisks dataset. An API can be considered as a processing step between user input and database software. The database portal's frontend (defined as what a user sees and interacts with on their browser) does not use hard-coded data: what is displayed is dynamically determined from the data available on the API. The frontend contrasts with the backend database portal services, such as the API and database software, which are not primarily designed for human interaction. 

Some website-building frameworks allow data to be transferred directly (and securely) to the frontend from a credentialed (i.e., password protected) database software. However, the SpExoDisks project splits the API into an independent service from the frontend, viewable at \url{spexodisks.com/api}. This breakpoint provides an opportunity to showcase available API data views as specific URL links on a static page. As a result, our frontend developers and users can visualize the live database portal's API from their browsers. The data provided by the SpExoDisks database portal's API is viewable in a compact format (such that there are no extra white spaces or new line characters) which can be beneficial for automated access from other websites and data pipelines. However, viewing the data across multiple lines and varying intention levels is helpful for human readers to see the data structure -- which is why the SpExoDisks API provides both data views.

The API can also process data from the frontend users and send it to the database software.
Take, for example, the need to store usernames and encrypted passwords -- which is required for data download in SpExoDisks -- in a MySQL database table. The API receives data when the frontend user submits a form and then processes that input into a format suitable for storage in the SpExoDisks MySQL database software. An overview of this process is also discussed in \S \ref{s.public-server} in the context of interactions between the services of the database portal.

The SpExoDisks team believes that astronomy data should be accessible to all. This means making our data available and viewable from web browsers and giving data context so that it can be easy to understand and interpret. While many programming language can read the JSON\footnote{\url{www.json.org}} data (often as key:value pairs, similar to a Python dictionary) from our API, we also provide a community tool for API access in Python (\url{https://github.com/spexod/spexod}), installable with Python's \texttt{pip} function.
Support for data science and machine learning is part of our database portal's API design.
In other words, the SpExoDisk server is inherently set up to perform computationally inexpensive operations that support machine learning questions (e.g., developers writing custom computations that occur directly in the MySQL database software). This design allows on-demand queries from users to stream our data directly from the API, making it easy to download the entire database as needed (anonymous users may download the entire database via API once a day). However, the fastest performance will always be when the database is on the same network as the machine learning process. Therefore, we recommend that users set up a local copy of our database and API, allowing further customization to their machine learning applications without network latency and throttling. We look forward to future partnerships where our data is downloaded automatically for multi-wavelength analysis, machine learning, or to display our data on another website.

\section{Data Interactivity and Accessibility}\label{s.data_interactivity_accessibility}

Astronomy data can be expensive to obtain, and repeated observations can require extensive justifications. As a result, the astronomy community believes observational and analysis data should (eventually) be posted to support \textit{access} by the astronomical community and general public -- where \textit{access} means that the data is made reachable or obtainable by the public. In \S \ref{s.begin-portal} and \ref{s.database_software}, we showed solutions for data access. And for the goals of some database portals, it is be enough to provide access via the API and/or to simply host files in a directory system that are available for download.

While the SpExoDisks database portal provides data access, our mission is also to provide an \textit{accessible} dataset to the community and public, where \textit{accessible} data is simple to retrieve but is also easy for anyone to understand. We wanted to create data visualizations to allow our experts and novice users to interactively examine spectra within the context of other astronomy data without downloading, importing, and displaying the data. 

\begin{figure*} 
    \centering
    \includegraphics[width=\linewidth]{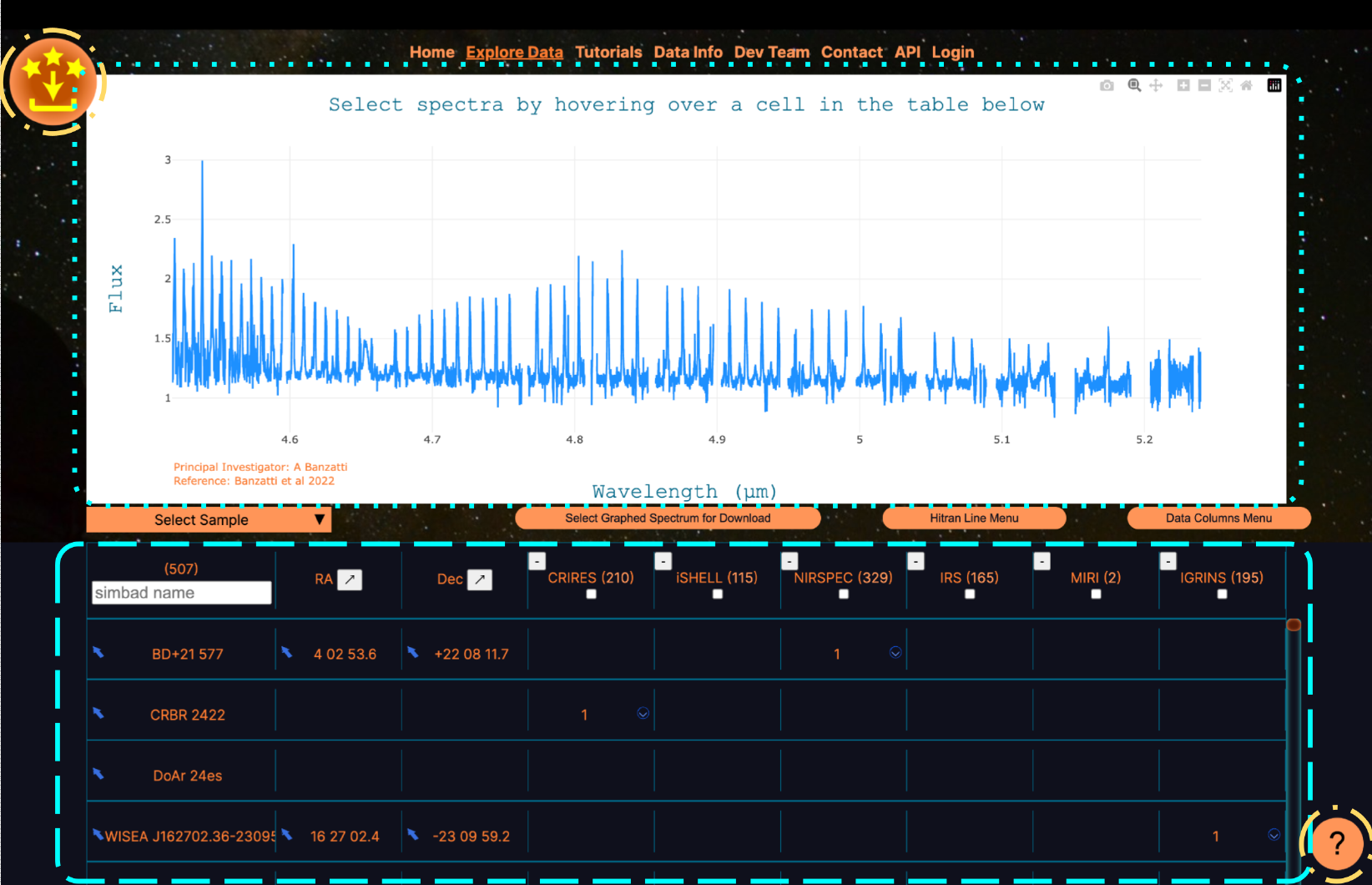}
    \caption{The SpExoDisks database portal as viewed from \url{spexodisks.com/ExploreData}. Four regions have been denoted with dashed and dotted highlights. The spectrum plot (shown in a dotted rectangular box) is an interactive plotting tool displaying spectra and molecular lines. In the bottom half of the figure, a long-dashed box indicates the navigation table, which is used for searching available stars and displaying associated spectra. The upper left corner has a long-dash-dotted circle showing the toggle for the spectra download menu. In the lower right, a short-dashed circle shows the help button, which toggles a window that explains the ``Explore Data" page of the SpExoDisks database portal.}\label{fig:anotated_explore_data}

\end{figure*} 

The ``Explore Data" page's spectrum plot (at the top of Fig. \ref{fig:anotated_explore_data}) allows experts to immediately assess the quality of the spectrum and inspect for features, such as molecular lines.
The navigation table (the long dashed box in Fig. \ref{fig:anotated_explore_data}) has a star name searching tool that evaluates several SIMBAD star names per object, allowing users to go directly to a spectrum of interest for a given star and instrument. The row of star data shows users what instruments and wavelength ranges are (and are not) currently available in our dataset, provided as columns. Each cell displays a number that indicates the available observations and a selection menu that provides a list of those observations in order by minimum wavelength value. Clicking on a spectrum link requests data from the API and updates the interactive spectrum plot. Users can view contextualized data to better understand what is available or what complimentary observations might yet be needed for a specific science case.
For non-expert users on the ``Explore Data" page, we provide a help button in the window's lower right (see Fig. \ref{fig:anotated_explore_data}) and indicate that certain features of the database portal may be interacted with by changing the color and brightness of the text when hovered over. 

Importantly, the SpExoDisks frontend developers continue to make updates for the large fraction of the world who have visual impairments. For example, we provide dynamic text sizing, which makes the website viewable for mobile users, small screens, and anyone who wants the website to display larger text sizes. Mouse-over text is available for many visual elements within the interactive plot, the navigation table, and menu buttons. We try to provide multiple options for each action and multiple ways to display information. 

Another common visual need is an accommodation for color blindness. The SpExoDisks frontend development team considered colors and color blindness from the beginning of the project. Within the frontend's codebase, we defined a narrow color pallet of five colors (per the Coolers\footnote{\url{https://coolors.co}} application) to be used in the rest of the application's components. By using a limited number of colors, we can easily ensure that each has a different hue and brightness level, so that each color will contrast with the others when used in any combination (tested using the Sim Daltonism\footnote{\url{https://michelf.ca/projects/sim-daltonism/}} application using a monochromatic filter and for a variety of color blindnesses). In addition, when viewing the graphed spectrum with molecular lines, the molecular lines of different isotopologues have not only varying colors, but we also changed the line styles between isotopologues to increase communication and understandability.

With all of this in mind, we have plans for additional improvements, such as designing an efficient tab-selection tool for the navigation table. Overall, we are open to expanding accessibility by looking for partnerships and making upgrades that would improve the SpExoDisks portal as an introductory teaching tool for spectroscopy for the scientific community and the public.

\subsection{Data Visualization}

Visualizations can be vital to understanding any data. Visualizations for a database portal share many of the same functionalities as figures within papers or proposals: they must concisely show a specific aspect or trend within a dataset. 
Science data visualization tools must also be accurate, displayed in standard formats and units, and referenced with literature sources. However, it was important for the goal of the SpExoDisks project that the database portal's spectrum plot (the dotted box in Fig. \ref{fig:anotated_explore_data}) must also be dynamic such that it will dynamically load data from the API and respond to user interactions. 

The SpExoDisks design process started by considering different hypothetical users. When designing the ``Explore Data" page's spectrum plot for an expert user, we wanted to present a dynamic version of a literature-quality figure in a format that is immediately recognizable and cites sources for the primary and context data. Expert users can view and explore trends within a familiar context, allowing the data to be incorporated into other publishable works. In addition, while sharing a publishable visualization is enough for static data, a dynamic website can go further. Each spectrum in the SpExoDisks database has an associated URL for users to share their favorite spectra with other collaborators. A fronted visualization system can dynamically display different types of data or other views of a dataset, such as zoom and panning options to show a spectrum's details (the dotted region in Fig. \ref{fig:anotated_explore_data}). Toggleable molecular lines offer additional context and user customization, viewable by clicking the ``HITRAN Line Menu" button (below the spectrum plot in Fig. \ref{fig:anotated_explore_data}). Users can then export the plot via built-in tools or taking screenshots to share the spectrum plot views. To facilitate citation to the original data, we display a fixed literature citation visible as users zoom on different spectral regions or when the plot is exported. 

For novice users, we wanted to consider different but compatible interactions with the ``Explore Data" page. Unlike the expert user, who may know how to get quickly to a specific aspect of a spectrum, a novice user may wish to have more support in learning how to explore and interpret the data. This means employing techniques that users already expect from website applications (i.e., mouse-over text, instruction pages, tutorial videos, help buttons, or friendly pop-up instructions) to aid in the understanding of what the database portal can display. However, too much information overwhelms new users, so we strove to make our default visualizations as simple as possible. Upon seeing a graphed spectrum for the first time, we wanted a novice user to have a less cluttered view of the presented data and provide many resources to explain the information displayed. We also tried to reduce the complexity of our interface by decreasing the steps needed to perform any action and by adding thoughtful defaults, such that users can select more complex and bespoke graph visualizations as they develop skills. In the SpExoDisks database portal, we hide our menus by default (such as the data download menu) to reduce the visual clutter and to direct users to focus on the spectrum plot and the navigation table (shown in Fig. \ref{fig:anotated_explore_data}). However, we show the download menu when the page loads for the first ten seconds to remind users we have a data download tool. 

Given resources and community interest, it is possible to continue extending the SpExoDisks database portal's visualizations to become an online laboratory for spectroscopy. The SpExoDisks team has considered extending the spectrum plot to allow users to process spectra by determining differential velocity shifts, tagging molecular lines and species, and overlaying models and other reference data. Another potential new feature could extend the plotting visualization system and allow users to upload and inspect their spectra using our existing tool set. And, after publication within the literature, perhaps authors and principal investigators could upload, verify, and process spectra into our database portal using the existing pipeline that uses the software environment as the database portal's API.

\subsection{Displaying Contextual Data}\label{s.display_context_data}

As discussed in \S \ref{s.data_context}, data needs context -- not only to be scientifically valid but also to make the data more useful. SpExoDisks enriches our spectral data with the contextual information related to the observed star, such as distance, effective temperature, or disk inclination angle -- which can be seen by adding columns to the navigational table in the ``Explore Data" page's bottom half in Fig. \ref{fig:anotated_explore_data}. This table can display and sort by all object parameters, which is possible since we require the parameter values to be displayed with a single unit, e.g., K, pc, string, or no-units.

Additional contextual data that SpExoDisks provides are molecular lines from the HITRAN database 2020 release \citep{hitran20}. A toolbar for toggling lines from different isotopologues of CO and H$_2$O allows users to identify these lines quickly and easily. The overlay of the molecular lines on the spectra conveys a huge amount of information that is most easily understandable in a graphical setting. For expert users, this may give a sense of the temperature of the gas emitting the observed spectrum or explain why a part of the spectrum is missing, in the case of H$_2$O lines for ground-based spectra. Because we have a rich dataset for each spectral line, we allow users to limit the lines displayed based on the line's upper and lower-level energy. All CO and H$_2$O lines between the spectrum's minimum and maximum wavelengths in the curated set are also provided as a part of the per spectrum data download (\S \ref{s.downloadable_content}). We note that at this time the downloaded spectral plots are 72 DPI and do not contain annotations of the spectral lines. Future upgrades may incorporate toggleable labeling for the molecular lines that include flexible annotations, as well as higher resolution downloads (e.g., 300 DPI) suitable for publication.

Incredibly important for finding spectra from a specific object, SpExoDisks provides a comprehensive set of star names, as discussed in \S \ref{s.data_organization} and Appendix \ref{s.star_name_problem}. The navigation table's name search box looks at a list of 30+ star-catalogs that are also available on SIMBAD, including Gaia (DR1/2/3), 2MASS, Henry Draper (HD), Tycho, Washington Double Star Catalog (WDS), and many others. Additionally, we allow custom ``common" names (e.g., Teegarden's Star) to be consistent with literature, popular culture, or the notation used for the original observations. The 30+ star name catalogs also provide a hint regarding other resources or data available for a given star. For example, Gaia DR3 names signify available properties from that catalog's massive dataset, while a WDS name indicates that an observed object has a companion or is a component of a multi-star system. 

When building the SpExoDisks database portal, we imagined a future where our users wanted to look at multiple spectra to make associations between multiple objects. As a result, we have abstracted the process of adding all our contextual data and are ready to partner with other to add more contextual data. With the speed at which users can stream our data, we are ready to support future data science that want to use parts of the SpExoDisks database portal as a sample for data science, modeling, and/or machine learning.

\subsection{Downloadable Content} \label{s.downloadable_content}

It is a core goal of the SpExoDisks database portal to provide downloadable content (e.g., spectra and contextual data) to scientists. However, this was one of the last major functionalities to be deployed as part of the database portal. Even with the database software and API in place, we found it more intuitive to design a parallel route in the data processing pipeline compared to using the existing views. The new route created files that can be found and downloaded based on API requests. This provided more flexibility compared to the alternative method to start from the existing API data views to assemble data and prepare files for download.

The SpExoDisks database software only has one data format (see \S \ref{s.database_software}) for each spectrum in the database portal. However, we process and present our data in two additional formats: human-readable UTF8 text files and astronomy-standard FITs format. When a user selects any number of spectra to download, both formats of files are located and copied to a zip file for download. The text and FITs formats are embedded with contextual metadata, including many known names, stellar object parameters, and literature references. The FITs files also include the original FITs header, when available. Two other files are also included with any download: 1) a README.md file explaining the format of the FITs files for those who are unaccustomed, and 2) a Python file that allows you to open and parse the FITs files and all the various types of contextual data. We included multiple formats, instructions, and Python code to provide accessibility and empower our users to extract and import our data.

The spectra download menu (shown in the upper left corner of Fig. \ref{fig:anotated_explore_data}) in the SpExoDisks database portal prompts users to log in and provide an email address. This is not meant to limit data access, but rather to provide a way for us to contact users about any major corrections or updates to the data. We found this to be necessary for the integrity of the data and to establish a chain of custody for all of the data we display. While we have yet to issue corrections to the data we provide, we feel confident that we can alert the community we serve to any errata or other issues, should the need arise.

\subsection{Programmatic Access}

Accessing our data via the terminal, within code/scripts, or in any other kind of programmatic manner is a form of accessibility. While frontend visualizations are helpful for understanding, assessing, and sharing data, they are not easily automated into processing pipelines. To be a resource for our user's processing pipelines and other astronomy database portals, we allow full access to all available data via a RESTful API, or REpresentational State Transfer API, which uses standard internet protocols that simplifies requests and responses. We deployed a Python package, \texttt{spexod}, which is installable via \texttt{pip}, that has an associated public GitHub (\url{github.com/spexod/spexod/}) and a documentation website (\url{spexod.github.io/spexod/index.html}) where users can see available functions for retrieving data from the SpExoDisks API. 

Users can interact with the SpExoDisks database portal's API directly from their browser at \url{spexodisks.com/api/}, which displays the URLs to all data, including spectra, per-object information, object name data, and all the viewable data from the front-end. Adding the suffix ``\texttt{?format=json}" to any route switches the data view from a human-readable HTML to a machine-readable JSON; this is how all data is consumed from the frontend browser application. And because the SpExoDisks database portal supports the ``GET" request method that can be used for retrieving data from RESTful APIs, many programs and programming languages have access to our data. Small data objects (small enough not to crash the browser) can also be copied and pasted from the browser into a Python script as a dictionary object. The SpExoDisks developers find these data views useful as a starting point for frontend feature development as well as another excellent break-point that allows human inspection of the processed pipeline data.

\section{Technical}\label{s.technical}

One of the most significant challenges to launching the SpExoDisks database portal was identifying the technologies and software environment scopes that were important to consider. We also researched and employed new systems as we progressed from data science pipeline to website development. After years of rewriting, redesigning, and updating, we use a set of technologies (software, programming languages, cloud hosting) and methods (how developers interact with the project) that are sustainable for our team. We present our technologies and methods to allow others to plan and compare projects based on our optimization from 2020 through 2023 for a specific science application. Unavoidably, the progress of software and process design will render this breakdown less relevant over time, but we hope that it may still serve as a useful starting point. However, updates will be visible through our public repository (\url{https://github.com/spexod/Portal}) which orchestrates all services and houses the code and configurations for NginX\footnote{\url{www.nginx.com/}}, MySQL, API, and the data processing pipeline.

Budget and time management are essential constraints that should be considered in database portal design for those who would like to provide public access to their data or data within their subfield.  To aid such individual considerations, we discuss the person-hours required to develop and maintain a codebase and documentation required to support a database portal similar to the SpExoDisks database portal: \\
$\bullet$ When hiring a developer for any new group, we recommend budgeting two years of salary to cover the time of a professional developer. This will give a research team time to organize their data and understand the product they want to present. However, once a team understands its database portal goals, data contextualization, data organization, data standarization and verification, and overall design (see \S \ref{s.begin-portal}), new projects can be deployed to the maintenance phase within a single year. \\
$\bullet$ During the time of database portal development (e.g, Year 1 for the SpExoDisks database portal), one hour of developer time is spent for every two hours of student time. This is because the leading developer spends a lot of time writing documentation, conducting surveys, and understanding research tools before developing new projects for student projects. Without existing examples, students start with more abstract tasks, which require more training.\\
$\bullet$ Maintaining an existing website (e.g., not adding new features or data) usually requires $\sim$80 hours/year of critical developer time. This is to maintain the overall health of the website, since certain components may be found to be vulnerable over time, important updates need to be handled, security practices change and require adaptations, software loses support, cookie preferences change, etc..\\
$\bullet$ Maintaining student work (e.g., after Year 4 for the SpExoDisks database portal), one hour of developer/manager time is typically needed for every five hours of student effort. This time is often equally split between refactoring the database portal to be consistent with best practices and the existing codebase and project management, since even the most talented students need attention to provide a high-quality product.

\subsection{Technology Used in SpExoDisks}\label{s.tech_used}

Table \ref{tab:tools} lists all the technologies recommended and used in the SpExoDisks projects, along with links for quick access. One of the most significant operational challenges for the SpExoDisks team was getting the software to work on every team member's computer with a variety of operating systems (OS) and architectures. In the early days, we constantly rewrote our installation documentation for three different OSs. However, this didn't solve all of the problems for the existing team and there were significant new issues when setting up incoming students every semester. Today, we have removed most of the environmental differences across our team's computers using the popular container system Docker\footnote{\url{www.docker.com}} to deploy all our code within the specific, required environment. A container is a way to set up an isolated, minimalist environment for software; containers do not need an entire OS, so they use resources relatively efficiently on the host machine. Docker containers solve two problems for our team: 1) We have a way to check that the website can be deployed in a local test network of Docker containers, then send those exact containers to be deployed on the live website host; and 2) We have divided the database portal processes into modular services that can be independently updated and tested. In addition, this second point has allowed us to continue using local installations (not in a container) for debugging and rapid development, with only minor differences compared to the stricter container environment.

\begin{table*}[t]
    \centering
    \begin{tabular}{l | r r}
    \hline
    Service Type & Deployed on SpExoDisk.com & Also Recommend \\ [0.5ex]
    \hline
    \hline
    Data Science Pipeline & \href{https://www.python.org/}{Python}, custom data classes & Your favorite software\\
    Database & \href{https://www.mysql.com/}{MySQL} & \href{https://www.postgresql.org/}{PostgreSQL}, \href{https://www.mongodb.com/}{MongoDB} \\
    Application Program Interface & \href{https://www.djangoproject.com/}{Django} framework (Python) & \href{https://flask.palletsprojects.com/en/3.0.x/}{Flask}, \href{https://expressjs.com/}{Express}, \href{https://rubyonrails.org/}{Rails}.\\
    Website & \href{https://nextjs.org/}{NEXT.js}-\href{https://react.dev/}{React} (\href{https://www.javascript.com/}{JavaScript}) & \href{https://astro.build/}{Astro}, \href{https://angular.io/}{Angular}, \href{https://jquery.com/}{jQuery} \\
    Web Server & \href{https://www.nginx.com/}{NginX} & \href{https://httpd.apache.org/}{Apache} \\
    Could Hosting & \href{https://aws.amazon.com/}{Amazon Web Services (AWS)} & \href{https://cloud.google.com/}{Google Cloud}, \href{https://azure.microsoft.com/}{Azure} \\
    Container Management & \href{https://docs.docker.com/compose/}{Docker compose} (\href{https://www.docker.com/}{Docker}) & \href{https://kubernetes.io/}{Kubernetes} \\
    Version control & \href{https://git-scm.com/}{git} & \\
    File and Docker repository & \href{https://github.com/}{GitHub} & \href{https://hub.docker.com/}{Docker Hub}, Docker only \\
    Editor/IDE & \href{https://www.jetbrains.com/webstorm/}{Webstorm} \& \href{https://www.jetbrains.com/pycharm/}{PyCharm} & \href{https://code.visualstudio.com/}{VScode} \\
    Graph/Data Display & \href{https://plotly.com/}{plotly} & \href{https://bokeh.org/}{bokeh} \\
    \hline
    \end{tabular}
    \caption{Technology systems used in the SpExoDisks project with respective hyperlinks. Compiled in 2024, things may have changed.}
    \label{tab:tools}
\end{table*}

Using the MySQL database software has meant data is always available quickly with built-in sorting and organization. The SpExoDisks team understood that this was a best practices approach, but we were surprised by the way that learning MySQL changed how we thought about data. Lessons from the SQL paradigm improved the data science pipeline and the frontend's data hydration (i.e. data loading and distribution to other components) because we would continually update our data structures/tables to make processes efficient. Database abstraction tools -- like the Python packages SQLAlchemy\footnote{\url{www.sqlalchemy.org}} and Django\footnote{\url{www.djangoproject.com}} -- allowed us to do column-wise operations, summing, and other basic functions using the optimized code in MySQL (or other database software) indirectly with Python. The column-wise functionality enables complex calculations on a relatively small virtual computer that hosts the SpExoDisks database portal.

JavaScript, rather than Python, was used for frontend website development, which significantly boosted the group's productivity. JavaScript required some new learning and adaptation for our team, but we realized it was simply the right tool for website development. JavaScript runs directly in the browser, unlike Python which requires JavaScript to handle dynamic and interactive features. JavaScript has also had a community concerned with web development for years and has grown to cover a range of web development scopes and use cases. As a result, finding junior developers or university students familiar with JavaScript and/or one of its many web development frameworks is much easier.

While Table \ref{tab:tools} may be helpful to those wishing to learn new software/systems or start a new project, the actual number of systems that a SpExoDisks team member employs is much larger. It's also important to note that the SpExoDisks team has observed that developer experience and software quality benefit the most when software has an active and growing community of users. Continued development and security updates have pushed us to continually and automatically update all packages and other environment components when deploying our software to the SpExoDisks database portal. While it can require extra work to keep all aspects of the project as up-to-date as possible during deployment, our team believes that it's important to be open to using new software and techniques. However, this also means that deprecations and vulnerabilities have to be addressed before each new deployment, such that processes are retired, code deleted, and unused software is removed. While different projects may require different solutions, our work flow has allowed us to gradually keep the project modern while accounting for necessary padding time for our team to address significant changes.

\subsection{Data Science Processing Pipeline}\label{s.data_flow}

\begin{figure*} 
    \centering
    \includegraphics[width=\linewidth]{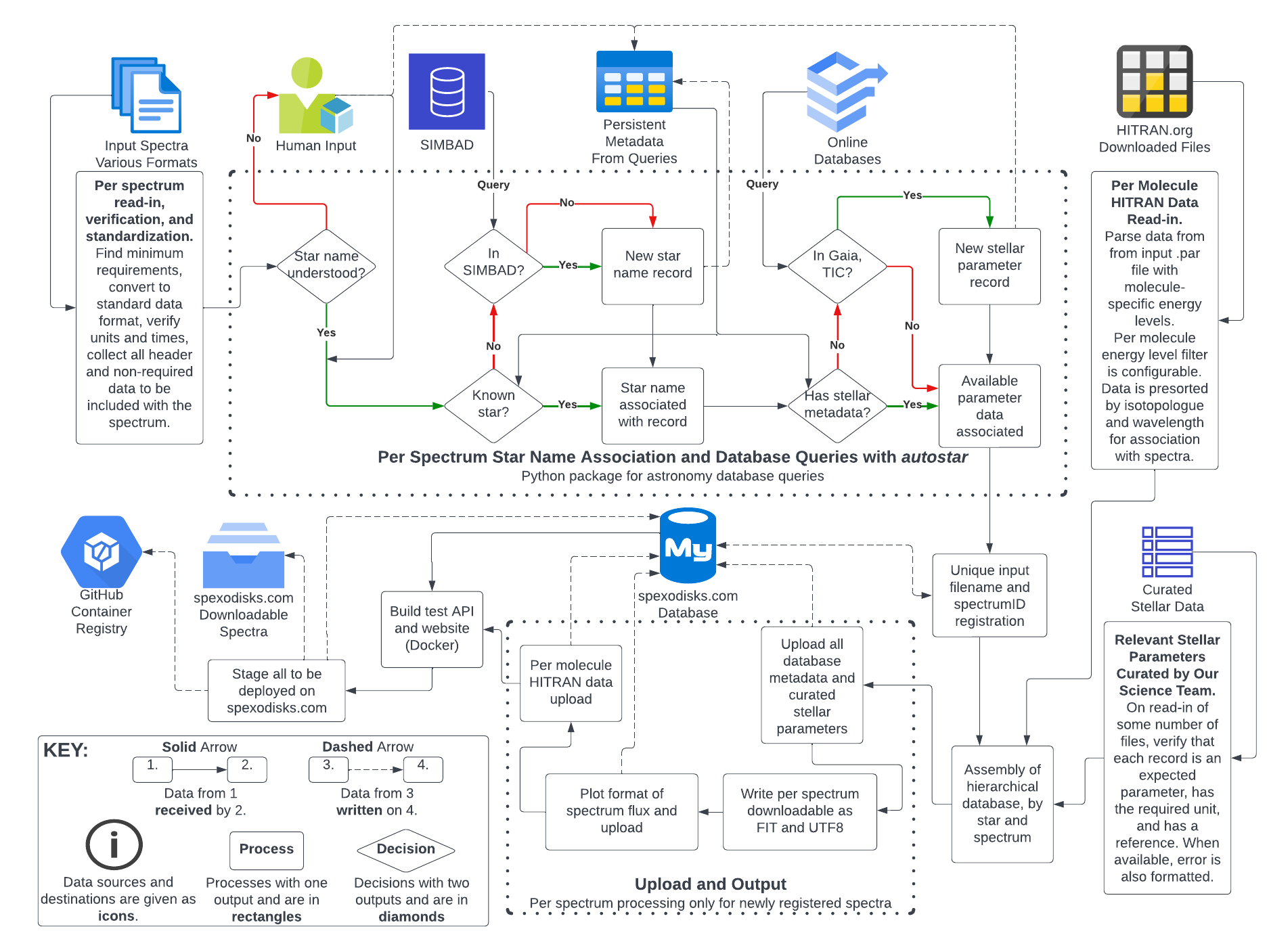}
    \caption{An organizational chart of the SpExoDisks data science pipeline (see the key in the lower-left corner). The pipeline starts at the top-left and processes clockwise, ending at center-left side of the diagram. The large dotted boxes denote pipeline process clusters that are triggered for each observed spectrum, where data read-in starts from icons that contribute data in the accumulation and context phase of the pipeline (top half to lower right). The second phase is public data export (bottom dotted box), which uploads data to a MySQL database software for efficient storage and retrieval. A locally-built, test website connected to the public server (``Build Test API and Website" process box) shows the newly staged data for inspection. The inspection environment is then configured to be deployed to the public website and uploaded to GitHub, completing the staging to update the SpExoDisks database portal.}
    \label{fig:dataflow}

\end{figure*}

The data science processing pipeline for the SpExoDisks database portal is a combination of two phases, which roughly corresponded to the top and bottom half of Fig. \ref{fig:dataflow}, respectively. First, an \textit{accumulation and context} system reads data products from various sources and links them to custom Python records (discussed more in Appendix \ref{s.custom_data_objects}). This is followed by a \textit{public data export} system optimized for delivering data to the SpExoDisks database software and API users. 

The SpExoDisks database portal has transitioned from primarily software development to a combination of development and maintenance. Starting at the data input icons in Fig. \ref{fig:dataflow} (top row, also the middle right), the accumulation and context system writes and accesses many files for intermediate processing pipeline data storage (e.g., SIMBAD, TIC, and Gaia); these files locally store the data retrieved from other astronomy database portals as plain text (e.g., CSV). With the benefit of hindsight, we now understand that our data processing pipeline would be faster and easier to maintain if we had used the same MySQL database software to save and access these intermediate processing results, instead of writing several custom micro-database software systems in Python. We have also made this pitfall in other areas, for example: by writing custom processes, we made systems more difficult for new students to learn and for our future selves to comprehend. In future data processing pipelines, we have resolved to use tools maintained by larger communities that continue to benefit from updates and documentation, enabling future change and maintenance.

All data objects in the SpExoDisks data processing pipeline are associated with a star name, with the exception of the HITRAN data. Per the discussions in \S \ref{s.data_organization} and Appendix \ref{s.star_name_problem}, data from any source must require a star name, triggering an association process (see the dotted box at the top of Fig. \ref{fig:dataflow}). The first time a new star name is encountered, the pipeline either activates an automatic data records retrieval from SIMBAD or prompts the user to create a new custom name that can be associated with a SIMBAD name or not. We save this name and other results from SIMBAD locally (\S \ref{s.data_context}).

While so much can be automated in the SpExoDisks data science pipeline, some things still need human attention. For example, receiving new spectra from any source requires negotiating formats and observation-specific conventions. This requires due diligence and double-checking by experienced and invested scientists (often the PI of the project). However, while new data may come with quirks, we have made tools that can report issues upon read-in. By strictly checking data at import, we can use custom errors and warnings that immediately point to relevant files, lines, and values found to be in an unexpected format. Sometimes correcting the data requires a direct change to the input file, while other times, a workaround can be added as a catch in the data processing pipeline. This step is taken for all data sources shown in Fig. \ref{fig:dataflow}. 

Once all the spectra are read-in and associated with a star, we register them with the SpExoDisks MySQL database software (see the last paragraph of \S \ref{s.data_organization}). The registration process allows us to upload only new spectra, as opposed to spectra that the database portal has already encountered. This cache is a crucial time-saving option as new spectra are observed at increasing rates (with improved spectral resolution), expanding our data storage requirements.

Spectroscopic line data from HITRAN requires a very different organizational structure (right side of Fig. \ref{fig:dataflow}). Within the SpExoDisks database portal, the HITRAN information is in tables that are organized by isotopologue and indexed by wavelength. The pipeline associates the molecular lines with a given spectrum as determined by the spectrum's wavelength range. This processing step also allows us to deliver all spectral lines available for each spectrum in the standard FITs file output provided for download at the SpExoDisks database portal's ``Explore Data" page, as mentioned in \S \ref{s.display_context_data}.

With all the data checked and assembled, data can be exported in the format required for the MySQL database software and API (bottom of Fig. \ref{fig:dataflow}). Metadata and curated stellar parameters are relatively small uploads ($\sim$10 MB) compared to all spectra ($\sim$10 GB), so metadata tables are updated each time and are allowed to change often. Next, each spectrum is outputted as downloadable files and uploaded to the MySQL database software in a format for the spectrum plot (Fig. \ref{fig:anotated_explore_data}). The only change in the plot format is the handling of null flux values where the plot only uses one null value per data gap (as opposed to one every wavelength interval), which can reduce the storage for some spectra by $\ge$20\%. We note that the downloadable data provides the null-fluxes as originally reported in the contributed spectrum, i.e. we do not make any changes from the source. As a final step, the relevant subsection of the HITRAN data is uploaded and updated as needed. All data is now staged for inspection and the data processing pipeline is completed.

We test a local version of the database portal using the staged data to examine how the updated data will appear on the public database portal. In this way we can do visual inspections using tools that are exactly the same as those on the public-facing content. In addition to inspecting, we export the environments used to view the test website as Docker images uploaded to the GitHub Container Registry (GHCR). With the Docker images exported and the downloadable files uploaded to the server, we can update the live website at-will with a script designed to have only 2-5 seconds of downtime for the public-facing database portal. While the API takes 30-45 seconds to restart fully, the frontend can use cached API data (see \S \ref{s.public-server}) from the image-build processes until the API service is available. 

We use the latest release of Docker images from Python and Node.js\footnote{\url{https://nodejs.org/}} as the basis for the SpExoDisks database portal's API and frontend services, respectively, to ensure we have the most current bug fixes and security patches. Staying up-to-date is important since updating the database portal will periodically fail to build a new image because packages and functions have become deprecated. While we try to address issues as they occur, we can always download the last working Docker image from the SpExoDisks GHCR account and proceed from that image until the issues preventing a stable, complete build can be adequately addressed.

\subsection{Public Server Configuration}\label{s.public-server}

\begin{figure}
    \centering
    \includegraphics[width=\linewidth]{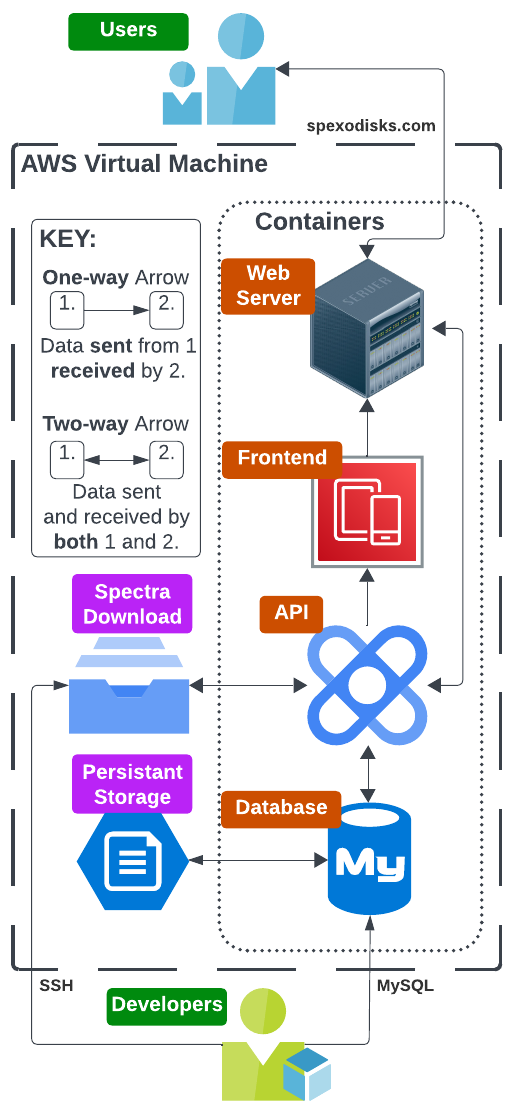}
    \caption{An organizational chart of the SpExoDisks public server configuration (see the key in the upper-left corner). Users (top-most icon) and developers (bottom-most icon) access the AWS virtual machine instance (icons enclosed in a dashed box) through specific network ports. Developers can update downloadable spectra files (filing cabinet icon) and the persistent storage for the MySQL database container (document icon), while users access the website from a browser. The Docker software allows four separate containers (icons enclosed in a dotted box) running software in specifically configured environments to be connected to external ports and each other. From top to bottom, the containers are a web server running NginX to route incoming URL requests, the frontend running Node.js to deliver cached API data with JavaScript and static files, and the API runs Python to shape and programatically access data from the fourth container running a MySQL database software server.}\label{fig:server_config}
\end{figure}

While the data science processing pipeline in \S \ref{s.data_flow} shows a changing progression, the public server configuration -- as shown in Fig. \ref{fig:server_config} -- should be considered an unchanging collection of software environments that work together to make the database portal. For SpExoDisks, the software environments are the Docker containers (located in the dotted box), where each container performs a service specific to a single software environment. Arrows show that data is transmitted between containers using standard network connections. The Amazon Web Services (or AWS) virtual machine (icons enclosed in the larger dashed box) hosts the Docker containers. The AWS server is relatively low-powered compared to most personal computers (with 4 GB Memory, 2 vCPUs, 80 GB SSD Disk with an IPV-4 address). The users (top icon) access the AWS virtual machine through their browser, while the SpExoDisks' developers (the bottom icon) manage the Docker containers, MySQL tables, and downloadable spectrum through specific network connections.

The four-container system is orchestrated to be all on the same network using the software Docker Compose\footnote{\url{docs.docker.com/compose/}}. The communication and ports are all configured in a \texttt{compose.yaml} file, which can then populate variables and authentication methods (i.e., passwords) to be shared with the containers to make secure run-time connections. Starting from an AWS virtual machine running Ubuntu, Docker is the only required software to install before being able to serve our website. As seen in Fig. \ref{fig:server_config}, database files for the MySQL container and the downloadable spectra are located in directories accessible to both the local file system and the relevant Docker containers. This creates persistent storage and backup for the MySQL data files and manages storage for updating the available downloadable files. Keeping the data separate from the containers' images helps to minimize image size and makes software updates independent of the data. Our images only provide the environment and our code. The authentication methods and data are only available when images are used as a basis to start a container instance.

Users interact with the SpExoDisks database portal via an HTTPS port monitored by a Docker container running an NginX web server. The web server does one task very well: it takes URL requests and sends them to the appropriate destination, which can either be static or dynamically generated. The URL routing can be complicated. For the SpExoDisks database portal, the URL requests with a suffix of ``/api/" are sent to the Docker container labeled ``API" in Fig. \ref{fig:server_config}, while all other requests go to the container labeled ``Frontend." Much of the metadata for the navigation table and home page is delivered from the ``Frontend" container at the same time as the HTML and JavaScript files. However, spectrum requests, HITRAN lines, and downloadable files are all accessed via API endpoints as users dynamically interact with the database portal.

Consider a timeline of events when a user visits the SpExoDisks database portal. Before the user navigates to our site, the frontend refreshes and caches the main data table, default spectrum, and other metadata from the API container. When the user visits \url{spexodisks.com}, they receive data from JavaScript, HTML, and other supporting files that create a browser application on the user's computer or mobile device. Specific user actions, like sorting the table, are implemented in JavaScript and done in the browser. Other actions, like plotting molecular lines and downloading spectra, require more data. In these cases, the browser sends a URL request to the API, which then handles and returns data in the formats expected by the browser application.  

Currently, the only user data saved in our database software is login information. That process starts with a form in the HTML and JavaScript browser application, which submits a username and password to the API from a URL request. The API checks to see if this is possible and returns a status that the browser application can interpret. If the status is ``success," a new record will be created in a MySQL database table. This record will be persistently available and will be independent of database portal updates (see ``Persistent Storage" in Fig. \ref{fig:server_config}).

\subsection{Database Software Access and Security}\label{s.database_access}
When choosing database software, it's important to consider how data will be updated and retrieved on the database portal. Fully featured database software like MySQL, PostgreSQL, MongoDB, etc. come with administrative tools for generating individually scoped credentials (e.g., username and password) for remote access. Directly reading HDF5 or CSV files, or using minimalistic tools on par with SQLite, will require teams to design remote data access/update processes for their files. However, individually scoped credentials and login capabilities are more relevant when the database portal is on cloud service, which often have limited computational resources compared to a server in an office or lab. There is no ``best method," instead it is a decision for individual teams to consider. The SpExoDisks database portal access is designed to accomplish three goals: 1) deliver read-access to all available SpExoDisks data for anyone, 2) allow specific team members to use database software to update the database portal, and 3) be able to delete then rebuild all services and access with one day of effort.

It could be unwise to publicly discuss the SpExoDisks team's policies for accessing and securing the database software. However, we can speak generally about our experiences working with fully featured database software. MySQL, MongoDB, and PostgreSQL can each be configured natively on a host system or in networked Docker containers and, with some effort, they can be set to require no password -- which is not advised. We recommend setting a very strong 50+ character root password on initialization and then creating non-root users for specific people and tasks. Current best practices recommend setting up a Secure Sockets Layer (SSL), a pair of keys that can encrypt data exchanged between two points on a network. SSL authentication is the default in MySQL 8.4, the release used by the SpExoDisks database portal at the time of writing. The first level of organization in the database software (confusingly called a \textit{database} PostgreSQL and \textit{schema} in MySQL and MongoDB) can often be scoped to specific user accounts. For example, while the API needs to access the top-level database that contains tables of user-data (or user-schema) for the website, a SpExoDisks team member working to upload new spectra data does not need access to the schema holding users login data. Narrowing scopes for specific actions or users is a best practice. We recommend saving credentials with limited access to a hidden file that can accompany any software needing a connection to the database portal.

If using Docker, it's a good practice to save credentials outside the image/container and pass them at runtime as environment variables. Usernames and passwords should not be saved in Docker images and they should be omitted from version control (e.g., git\footnote{\url{https://git-scm.com/}}). One advantage of building database software and other services in Docker is that your project can be developed independent of the OS, as discussed in \S \ref{s.tech_used}. The only OS-dependent code in the SpExoDisks project is a script that installs and configures Docker for Ubuntu on the AWS virtual machine. This makes the project extremely portable to other servers, both cloud and physical computers, and makes us less vulnerable to mistakes or ransomware attacks. While this is not true for all Docker base images (e.g., original or starting images on which subsequent images can be built), the base images used by SpExoDisks are available on several CPU architectures: x64 for CPUs from Intel and AMD, the arm64 tested on both RaspberryPi-4, and latest generation of Apple computers. 

Another benefit of fully featured database software is to provide data access to individuals outside the development team. For example, the Gaia collaboration allows uncredentialed users (e.g., public/anonymous without authentication or login) to write and submit database queries; as a result, this feature is used by the SpExoDisks data processing pipeline to automatically associate spectra with relevant Gaia stellar parameters. During development, the SpExoDisks team experimented with direct access to our MySQL database software that was both credentialed and uncredentialed before we had a complete vision of what our API should offer. We ultimately decided that only the API and the data science pipeline tools should directly access the MySQL database software (see Fig. \ref{fig:server_config}). While this provides an extra layer of isolation for the database software, the choice was primarily based on the API layer allowing additional data-shaping configurations and access to Python libraries. This is further simplified for our team, because the API only writes to the database portal in certain circumstances, such as adding and storing encrypted user data or resetting the login credentials. 

We note access to the SpExoDisks database portal could change in a future upgrade to allow designated users, such as PIs, to upload new spectra and data descriptions on their own. While we are interested in developing this kind of portal, it would require additional collaboration and support. Therefore, at this point we recommend that those interested in providing spectra contact us directly.

\section{Science Application of the SpExoDisks Data}\label{s.science_application}
\begin{figure*}
    \centering
    \includegraphics[width=\linewidth]{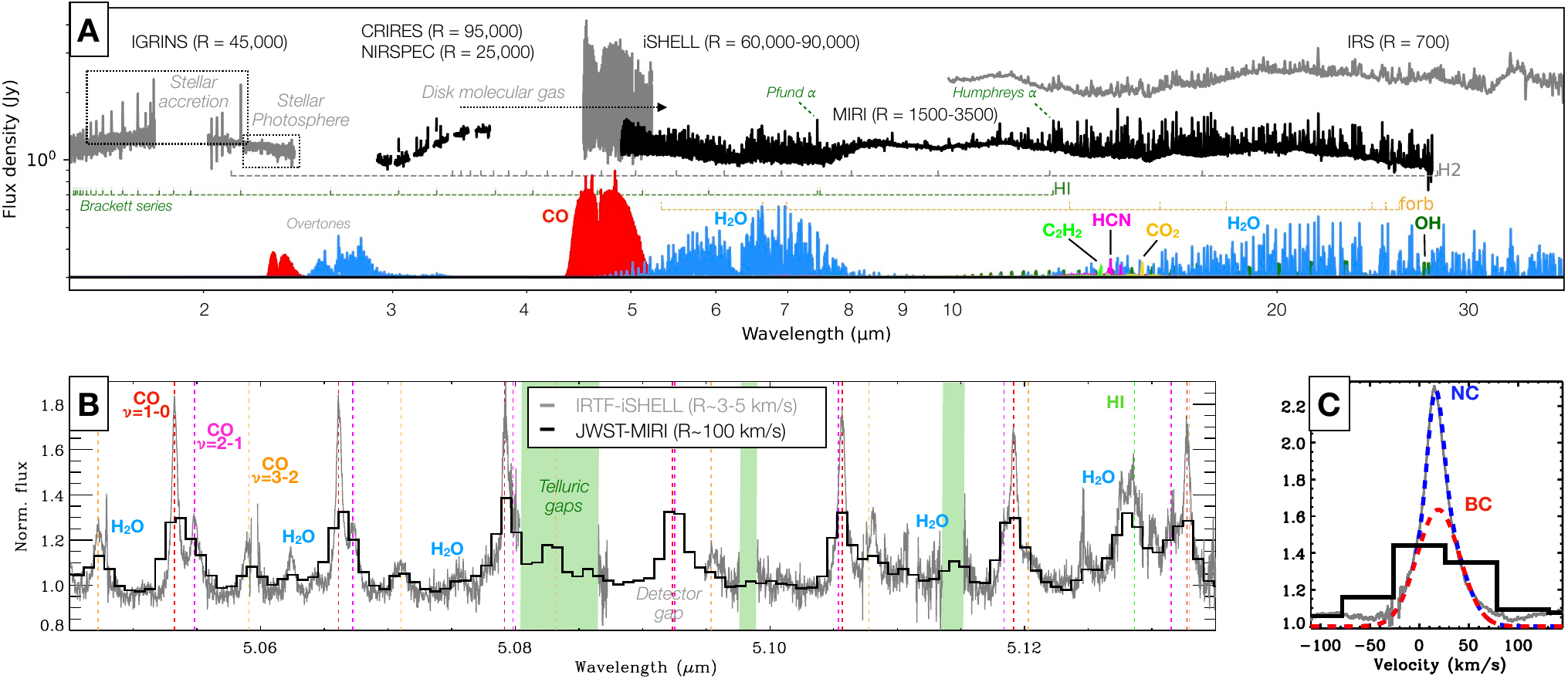}
         \caption{\textbf{A}: Overview of spectra included in SpExoDisks, using examples and similar plotting tools to what is included on \url{spexodisks.com}. Different instruments and their resolving power are labeled. Synthetic models of the main molecular species are shown at the bottom for reference. Regions of astrophysical interest (e.g., stellar accretion and photosphere, disk molecular gas) are indicated in gray. \textbf{B}: Zoomed-in portion near 5~$\mu$m, illustrating CO and H$_2$O emission spectra of the protoplanetary disk in FZ~Tau as observed with iSHELL \citep{banzatti22,banz23} and MIRI-MRS \citep{pontoppidan24}; the lower resolution of MIRI causes the blending of different transitions and species that are observed with iSHELL (labeled in different colors). \textbf{C}: Zoomed-in visualization in velocity space of an individual CO $\nu = 1-0$ line profile as decomposed into two kinematic components, a broad component (BC) in red and a narrow component (NC) in blue. The broadening of these kinematic components informs on temperature and density gradients in the inner disk surface and wind.}\label{fig:spectrum_ex}
\end{figure*}

The most unique and powerful scientific value of SpExoDisks is the combination of spectra with different resolving power and wavelengths (either or both), which enables to reach goals that individual instruments and datasets alone cannot provide. In this section, we describe a brief example in the context of the new spectra of protoplanetary disks that are being observed with the James Webb Space Telescope (JWST) and published at the time of writing this paper \citep[e.g.][]{grant23,banz23b,ramireztannus23,pontoppidan24,Temmink24}.

Fig. \ref{fig:spectrum_ex} provides an overview of spectra currently included in SpExoDisks, to illustrate their overlap or complementarity in spectral coverage and resolving power (R = $\lambda/\Delta\lambda$). The near-IR region is covered from the ground at high resolving power with the following instruments: H band ($\sim$1.3-2.0$\mu$m) and K band ($\sim$2.0-2.4$\mu$m) with IGRINS \citep[R~$\approx45,000$,][]{igrins}, L and M bands with CRIRES \citep[R~$\approx80,000-95,000$,][]{crires}, NIRSPEC \citep[R~$\approx25,000$,][]{nirspec}, and iSHELL \citep[R~$\approx60,000-90,000$,][]{ishell}. The mid-IR region is covered from the ground with VISIR \citep[R~$\approx30,000$ but only on 0.1~$\mu$m-wide spectral settings,][]{visir}, and from space with IRS before \citep[R~$\approx700$,][]{irs} and now MIRI-MRS \citep[R~$\approx1500-3700$,][]{miri,miri2,jwst-mrs,Argyriou23,pontoppidan24}.

\begin{figure*}
    \centering
    \includegraphics[width=0.8\linewidth]{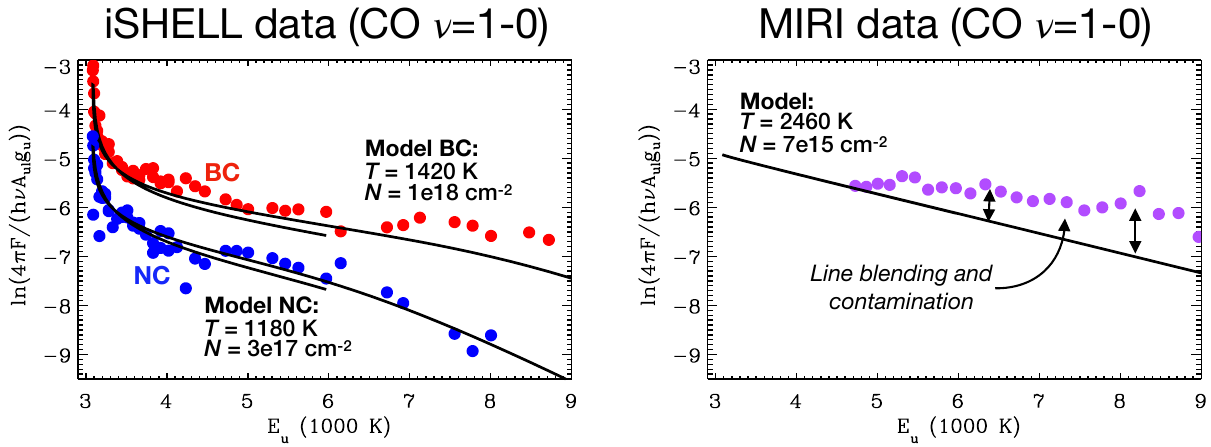}
         \caption{Left: Different excitation of the two kinematic components shown in Fig. \ref{fig:spectrum_ex} (panel C) as visualized in the population diagram using line fluxes extracted from the SpExoDisks spectra. Single-temperature models are overplotted for comparison, with best-fit values shown in the figure. Right: The two components are blended at the lower resolution of MIRI (Fig. \ref{fig:spectrum_ex}, panel C), and only the higher $J$ levels at $> 4.9 \mu$m are covered in the spectrum (Fig. \ref{fig:spectrum_ex}, panel A); as a consequence, the best-fit model only finds a very high temperature as mimicked by the flat part of the rotation diagram. }\label{fig:res_temp}
\end{figure*}

The global spectral coverage obtained by the combination of multiple instruments is extremely rich in tracers of gas within the system (Fig. \ref{fig:spectrum_ex}). For example, the spectra show lines from the stellar surface (absorption lines in the H and K bands on the left) to the accretion region (HI lines from the Brackett series in the H band all the way to the Pfund and Humphreys series; see the $\alpha$ line of each series in Fig. \ref{fig:spectrum_ex}). In addition, the inner disk winds and outflows are observed through the \ce{H2} and forbidden lines from \ce{Ne} and \ce{Fe}, while the inner disk chemistry emits a dense forest of molecular lines from ro-vibrational and rotational bands from \ce{CO, H2O, OH, HCN, C2H2, CO2} and some of their isotopologues. 

The lower panel of Fig. \ref{fig:spectrum_ex}, panel B, zooms-in on a region where space- and ground-based spectrographs overlap. The comparison of spectra between MIRI-MRS on JWST and iSHELL on IRTF in this region demonstrates the advantage of combining multiple instruments that observe the same spectral region. For example, the very high resolving power obtained from the ground with iSHELL reveals the gas kinematics in high detail by spectrally resolving the velocity profile of individual emission lines. On the other hand, space observations from MIRI-MRS lose the kinematic detail but gain sensitivity and also fill in telluric gaps not accessible from the ground (see labeled examples) to provide complete spectral coverage. Panel C zooms-in even more to show the velocity profile of an individual CO line. This illustrates two kinematic components (called broad or BC and narrow or NC) that can be extracted from the spectra which trace gas at different radial distances (i.e. a different Doppler broadening from gas at a different Keplerian velocity) from the star \citep[e.g.][]{bast11,bp15}. These high-resolving-power spectra also show the distinction between Keplerian disk (which have the characteristic double-peak profile) and disk+wind line profiles \citep[which have a narrow single-peak line center; for more examples of different line profiles see][]{banzatti22}.

Fig. \ref{fig:res_temp} demonstrates the scientific outcome of combining CO spectra from different instruments illustrated in Fig. \ref{fig:spectrum_ex}, panel B. Here we use a population diagram to illustrate the excitation of CO, a convenient diagram widely used in astrophysics to study molecular gas emission \citep[e.g.][]{popdiagram}. In this diagram, a different temperature of the gas produces a different slope. In addition, a variation in gas column density increases the curvature of the distribution of observed line fluxes as a function of their upper level energy, $E_u$. The excitation of the two distinct components from Fig. \ref{fig:spectrum_ex}, Panel C, becomes very visible in Fig. \ref{fig:res_temp}, left, from their different slope and curvature. 

By fitting the observed line fluxes with a plane-parallel ``slab" model of gas in the disk, that includes temperature (T) and column density (N) as free parameters \citep{iSLAT}, we can estimate these property differences in the protoplanetary disk of FZ~Tau -- provided as example in Fig. \ref{fig:res_temp}. These fits show that the two BC and NC components have a factor $\sim$3 different column density and different rotational excitation, where NC has $T \sim 1200~K$ and BC has $T \sim 1400~K$ \citep[the difference between the two components can be even larger, as found in other disks in][]{banzatti22}. These individual models show us the properties of a disk surface where temperature and density decrease with radius, from a hotter denser inner region closer to the star ($\le$ 0.1~AU) out to colder CO gas at larger radii (1--10~AU), covering the entire planet-forming region where exoplanet populations are observed today. The possibility to extract these properties as a function of disk radius is fundamental to understand the physical and chemical evolution of inner disk regions that are not accessible with direct imaging \citep[see recent overviews in][]{banzatti22,banz23}.

On the other hand, the lower resolution of MIRI-MRS spectra blends different kinematic components together (as we have seen in Fig. \ref{fig:spectrum_ex}, panels B and C), and covers only the high-$J$ region of the ro-vibrational band of CO ($> 4.9 \mu$m), therefore missing the curvature produced by high optical depth at lower $J$ levels (Fig. \ref{fig:res_temp}, right). The consequence is that a simple fit to the MIRI data will be very degenerate between optically thin and thick solutions (i.e. a low or high column density, respectively), which will then overestimate the temperature of the emission (due to the flattening in the curve from line blending, as indicated in Fig. \ref{fig:res_temp}, right).
All the JWST disk programs in Cycle 1 and beyond will need to use high-resolution ground-based CO spectra to support the analysis of space spectra and obtain gas emitting regions from the resolved kinematics to distinguish between fundamental scenarios in the interpretation of MIRI spectra, for instance the presence of small inner disk cavities that are beyond the reach even of ALMA \citep[e.g.][]{grant23,Temmink24}.

\section*{acknowledgments}
All authors would like to thank the entire development team (\url{spexodisks.com/DevTeam}), but specifically the students from Texas State University that brought the excitement, talent, and knowledge needed to explore new technology methods. For the climbers, may you never run out of mountains to climb.
The research shown here acknowledges use of the Hypatia Catalog Database, an online compilation of stellar abundance data as described in \citep{Hinkel14}. NRH would like to thank Tatertot and Lasagna for their help and support.
This research has made use of the SIMBAD database, operated at CDS, Strasbourg, France 2000, A$\&$AS, 143, 9 \citep{Wenger2000}.
This work has made use of data from the European Space Agency (ESA) mission
{\it Gaia} (\url{https://www.cosmos.esa.int/Gaia}), processed by the {\it Gaia}
Data Processing and Analysis Consortium (DPAC,
\url{https://www.cosmos.esa.int/web/Gaia/dpac/consortium}). Funding for the DPAC
has been provided by national institutions, in particular the institutions
participating in the {\it Gaia} Multilateral Agreement.
This work uses data from the TESS Input Catalog (TIC), \cite{StassunOelkers_2019} is a catalog of stars observed by the TESS mission.
This research has made use of ESASky, developed by the ESAC Science Data Centre (ESDC) team and maintained alongside other ESA science mission's archives at ESA's European Space Astronomy Centre (ESAC, Madrid, Spain), \cite{Giordano_2018} and \cite{Baines_2016}.
SpExoDisk's uses HITRAN (high-resolution transmission molecular absorption database) to display molecular line locations, see \cite{hitran20}.
This work includes observations made with the NASA/ESA/CSA James Webb Space Telescope. The data were obtained from the Mikulski Archive for Space Telescopes at the Space Telescope Science Institute, which is operated by the Association of Universities for Research in Astronomy, Inc., under NASA contract NAS 5-03127 for JWST. These observations are associated with Cycle 1 GO program 1549.

\appendix
\section{The Star Name Problem}\label{s.star_name_problem}
Uniting an individual observation to all other data regarding the same star is essential for data contextualization in astronomy. Unfortunately, stars have many names (sometimes $>$50!). The brightest stars can often pose the most significant challenge as their names can be colloquial strings or Greek constellation references. These names often lend themselves to observational shorthand, inconsistent capitalization, varying blank spaces, and other transformations that are easy for a human to read and correct but are challenging to add to a database without human intervention. This presents a barrier to introducing and sharing data across various sub-disciplines and generations of astronomers. Consequentially, this barrier can be insurmountable for those outside of astronomy, who may not realize that stars may be referred to via different names. For the stars in the SpExoDisks database portal, many hours of collective effort have been saved using the object names provided by SIMBAD -- which excels at identifying star names. 

The SpExoDisks database portal places a high value on all spectra that have been generously contributed by collaborators for display and download. We work hard to identify star names in any format and confirm they can link to existing records. The core of our star name identification process is to find a name recognizable to SIMBAD and then use SIMBAD’s API to return the additional star identifiers. These identifiers allow us to access and query data from additional online databases and link objects identified by different star names. SIMBAD's API also provides the raw data that enables users of the SpExoDisks data portal to search for objects using many star names associated with a single star (as discussed in \S \ref{s.display_context_data}).

For the SpExoDisks database portal, we approach the star name as something that can be determined by parsing names within each imported stellar catalog. Even with excellent tools provided by SIMBAD, converting names into unique identifiers for keys in a database table can be complicated by variations in notation. Consider a star identified as BD+19 00706, using a reference to the Bonner Durchmusterung (BD) catalog \citep{BD1903}. The first step when parsing in SpExoDisks is always to make the string name lowercase to remove case-dependent matching. Next, the stellar name catalog is identified with the ``bd" prefix, which sends the remaining part of the string to a parser specific to the ``BD" names. The ``BD" parser first looks for a ``+’’ or ``-’’ character; however, since many observers often omit the ``+’’ symbol as implicitly defined, we must interpret a lack of any symbol to indicate ``+’’. With this accomplished, we must now parse two integer values. However, major issues are created when names are reported without leading zero, for example BD+19 706, since the number of expected characters is variable. To solve this issue for most cases, we convert both coordinate strings (before and after the blank space) to integer values. However, when a blank space is not present (for example BD+12346), it's unclear whether there should be a leading zero (BD+01 2346 -- an F7 star) or a trailing zero (BD+12 3460 -- a K5 star); in these cases, the star names need to be manually cross-referenced with other verifiable properties (such as RA and Dec) to confirm the correct name. Finally, the parser delivers a Python tuple (`bd’, `+’, 19, 706), which is a unique projection of the name information that can be compared to various input formats of Bonner Durchmusterung star names. 

Carefully parsing star names helps create unique data record keys within the SpExoDisks data portal. However, for many purposes, we also want the star name to be converted back into a string that is as close to a SIMBAD-friendly name as possible. We use stellar catalog (e.g., BD, TYC, 2MASS, etc.) specific definitions to convert the parsed tuple to a SIMBAD format that includes the capital letters, symbols, leading zeros, and blank space separators.

Even with substantial parsing and logic catered to individual stellar catalog name types, some notations for observed stars can still fail to be correctly parsed by the \texttt{autostar} Python package. In cases such as this, where a name cannot be parsed but may also not be incorrect, the star name parser can be supplemented with a lookup table converting popular/colloquial names to those recognized by SIMBAD. The lookup table is a simple, two-column CSV file allowing developers to add names manually. In addition, a built-in interface prompts users running the SpExoDisks data science code to enter a SIMBAD-recognized name when unparseable names are encountered. This is one of the many ways we have streamlined the new data induction process for the SpExoDisks project.

When selecting the default name for a given star on the database portal, or the database key, we established a preferential list of star name types to determine which name should have priority. In this way, we can choose the formatted name for an object with the highest available preference and provide a consistent database key for a given object between runs within the SpExoDisks data science code. However, a star may be known by multiple names within the same SIMBAD stellar catalog, for example, ** Coo 271a is known by both HD109573 and HD109573A. In these situations, we always pick the longer of the two names to represent the database key, since the longer name is likely to contain more information (e.g., that this system is a binary) that could be useful for our developers or users. For similar reasons, the WDS \citep{WDS01} catalog is the preferred naming system specifically for multi-star systems, because it offers important information about a member object’s status in a hierarchical multi-object system. If the names are the same length, we choose the name that comes first alphabetically. 

A consistent, single, unique string that matches the SIMBAD star name is an excellent database table key for identifying a star. However, these strings can contain special characters and blank spaces, making them non-optimal for use as Python attributes, MySQL table names, or URL strings. Therefore, a final, reversible step is required to convert special characters into a descriptive string of allowed characters, e.g., `*' becomes `star' and `$[$' becomes `leftbracket' within the star name. 

We note that updates are often required to process new stellar catalog names, especially as exoplanets are discovered and their names have not yet propagated to SIMBAD. An alternative method to simplifying star name searching/comparison is to remove stellar catalog-specific parsing and allow any name strings with blank space removed. This technique is tolerant to mistakes by expiring the data periodically, e.g., once a year, and then automatically updating names by rechecking SIMBAD. While not currently part of the SpExoDisks data science processing pipeline, this updated feature has been propagated through the Hypatia Catalog database portal and will eventually be included in SpExoDisks, as part of their shared approaches and strategies.

\section{Custom Data Objects}\label{s.custom_data_objects}

Working with data in Python allows us to make custom data objects that suit specific needs. While data is often expected to be in specific formats, there are usually a variety of caveats or circumstances that require special handling. Let us consider a data table, a familiar and straightforward data object. We often see tables with a single header row where each value in the header row indicates the type of values that will be found in that column, while additional rows of data will represent single items. Especially in astronomy, the first column is expected to contain a unique identifier that can name a row of data as a unique key. This simple example has a few rules and conventions. However, sometimes there are multiple rows with the data for the same star (either intentionally or because different names were used for that star), possibly with conflicting values of data within the columns. Perhaps the table was edited in a text editor that left hidden characters (e.g., carriage returns) that breaks the parsing routine or changed the standard conventions. The SpExoDisks project uses many tables as raw input for the data processing pipeline, and we have experienced a lot of variation.

To check for various non-standard table possibilities, we can add a validation step in the key-value assignment stage of a custom dictionary object to check data as it is assigned during read-in. This can help identify mistakes faster and more directly while adding new data. Data validation can be accomplished by extending existing Python classes to do the required checking, such as a dictionary that forces all keys to be lowercase strings or a complex check that enforces units and references. When possible, such tools can solve problems; but for issues that cannot be resolved, we raise errors are called `exceptions.' 

Python makes it simple to subclass existing objects like lists, tuples, and dictionaries (practically, NameTuple and UserDict are preferable to Tuple and Dict) and then add new custom methods that override methods in the superclass. Shown in the example below (also posted at \url{https://github.com/spexod/Portal/blob/main/backend/examples/user_dict.py}) is a Python class with instances of data objects that do not allow duplicate keys. The \texttt{TableDict} object only allows strings to be used as keys (comment \texttt{\# 1}), and if it is not a string then a helpful error message is raised. To reduce ambiguity, any strings are converted to lowercase versions (comment \texttt{\# 2}). Finally, this custom dictionary object does not allow overwriting of existing keys (comment \texttt{\# 3}), which enforces that the data table must have a unique key for each row of data.

\begin{verbatim}
from collections import UserDict
class TableDict(UserDict):
    def __setitem__(self, key: str, value):
        # for setting key, value pairs as
        # a_table_dict[key] = value
        if not isinstance(key, str):
            # 1
            raise ValueError('TableDict ' +
                'only allows string keys, ' +
                f'got {key}, type {type(key)}')
        # 2
        key = key.lower()
        if self.__contains__(key):
            # 3
            raise KeyError(f'Key: {key}, ' +
                'already exists, repeated ' +
                'assignment is not allowed ' +
                'in TableDict')
        else:
            self.data[key] = value
\end{verbatim}

Building custom data objects that enforce certain rules or data topologies allows us to sustainably add data over long periods of time. Data is often written and read for the first time by an expert developer with a total understanding of the subject matter. However, subsequent updates are frequently made by student developers, outside collaborators, or even the experts themselves who forgot the intended rules in the months/years since the data processing pipeline was first written. These reasons exemplify why it's vital to enforce the intended logic of all data added to any database portal, so that critical error checking is provided to help developers find and resolve any conflicts. 


\begin{thebibliography}{}
\expandafter\ifx\csname natexlab\endcsname\relax\def\natexlab#1{#1}\fi
\providecommand{\url}[1]{\href{#1}{#1}}
\providecommand{\dodoi}[1]{doi:~\href{http://doi.org/#1}{\nolinkurl{#1}}}
\providecommand{\doeprint}[1]{\href{http://ascl.net/#1}{\nolinkurl{http://ascl.net/#1}}}
\providecommand{\doarXiv}[1]{\href{https://arxiv.org/abs/#1}{\nolinkurl{https://arxiv.org/abs/#1}}}

\bibitem[{{Argelander}(1903)}]{BD1903}
{Argelander}, F. W.~A. 1903, Eds Marcus and Weber's Verlag, 0

\bibitem[{{Argyriou} {et~al.}(2023){Argyriou}, {Glasse}, {Law}, {Labiano},
  {{\'A}lvarez-M{\'a}rquez}, {Patapis}, {Kavanagh}, {Gasman}, {Mueller},
  {Larson}, {Vandenbussche}, {Glauser}, {Royer}, {Dicken}, {Harkett},
  {Sargent}, {Engesser}, {Jones}, {Kendrew}, {Noriega-Crespo}, {Brandl},
  {Rieke}, {Wright}, {Lee}, \& {Wells}}]{Argyriou23}
{Argyriou}, I., {Glasse}, A., {Law}, D.~R., {et~al.} 2023, \aap, 675, A111,
  \dodoi{10.1051/0004-6361/202346489}

\bibitem[{{Astropy Collaboration} {et~al.}(2022){Astropy Collaboration},
  {Price-Whelan}, {Lim}, {Earl}, {Starkman}, {Bradley}, {Shupe}, {Patil},
  {Corrales}, {Brasseur}, {N{\"o}the}, {Donath}, {Tollerud}, {Morris},
  {Ginsburg}, {Vaher}, {Weaver}, {Tocknell}, {Jamieson}, {van Kerkwijk},
  {Robitaille}, {Merry}, {Bachetti}, {G{\"u}nther}, {Aldcroft},
  {Alvarado-Montes}, {Archibald}, {B{\'o}di}, {Bapat}, {Barentsen},
  {Baz{\'a}n}, {Biswas}, {Boquien}, {Burke}, {Cara}, {Cara}, {Conroy},
  {Conseil}, {Craig}, {Cross}, {Cruz}, {D'Eugenio}, {Dencheva}, {Devillepoix},
  {Dietrich}, {Eigenbrot}, {Erben}, {Ferreira}, {Foreman-Mackey}, {Fox},
  {Freij}, {Garg}, {Geda}, {Glattly}, {Gondhalekar}, {Gordon}, {Grant},
  {Greenfield}, {Groener}, {Guest}, {Gurovich}, {Handberg}, {Hart},
  {Hatfield-Dodds}, {Homeier}, {Hosseinzadeh}, {Jenness}, {Jones}, {Joseph},
  {Kalmbach}, {Karamehmetoglu}, {Ka{\l}uszy{\'n}ski}, {Kelley}, {Kern},
  {Kerzendorf}, {Koch}, {Kulumani}, {Lee}, {Ly}, {Ma}, {MacBride}, {Maljaars},
  {Muna}, {Murphy}, {Norman}, {O'Steen}, {Oman}, {Pacifici}, {Pascual},
  {Pascual-Granado}, {Patil}, {Perren}, {Pickering}, {Rastogi}, {Roulston},
  {Ryan}, {Rykoff}, {Sabater}, {Sakurikar}, {Salgado}, {Sanghi}, {Saunders},
  {Savchenko}, {Schwardt}, {Seifert-Eckert}, {Shih}, {Jain}, {Shukla}, {Sick},
  {Simpson}, {Singanamalla}, {Singer}, {Singhal}, {Sinha}, {Sip{\H{o}}cz},
  {Spitler}, {Stansby}, {Streicher}, {{\v{S}}umak}, {Swinbank}, {Taranu},
  {Tewary}, {Tremblay}, {de Val-Borro}, {Van Kooten}, {Vasovi{\'c}}, {Verma},
  {de Miranda Cardoso}, {Williams}, {Wilson}, {Winkel}, {Wood-Vasey}, {Xue},
  {Yoachim}, {Zhang}, {Zonca}, \& {Astropy Project Contributors}}]{Astropy22}
{Astropy Collaboration}, {Price-Whelan}, A.~M., {Lim}, P.~L., {et~al.} 2022,
  \apj, 935, 167, \dodoi{10.3847/1538-4357/ac7c74}

\bibitem[{Baines {et~al.}(2016)Baines, Giordano, Racero, Salgado, Martí,
  Merín, Sarmiento, Gutiérrez, Landaluce, León, Teodoro, González, Nieto,
  Segovia, Pollock, Rosa, Arviset, Lennon, O’Mullane, \&
  Marchi}]{Baines_2016}
Baines, D., Giordano, F., Racero, E., {et~al.} 2016, Publications of the
  Astronomical Society of the Pacific, 129, 028001,
  \dodoi{10.1088/1538-3873/129/972/028001}

\bibitem[{{Banzatti} \& {Pontoppidan}(2015)}]{bp15}
{Banzatti}, A., \& {Pontoppidan}, K.~M. 2015, \apj, 809, 167,
  \dodoi{10.1088/0004-637X/809/2/167}

\bibitem[{{Banzatti} {et~al.}(2022){Banzatti}, {Abernathy}, {Brittain},
  {Bosman}, {Pontoppidan}, {Boogert}, {Jensen}, {Carr}, {Najita}, {Grant},
  {Sigler}, {Sanchez}, {Kern}, \& {Rayner}}]{banzatti22}
{Banzatti}, A., {Abernathy}, K.~M., {Brittain}, S., {et~al.} 2022, \aj, 163,
  174, \dodoi{10.3847/1538-3881/ac52f0}

\bibitem[{{Banzatti} {et~al.}(2023{\natexlab{a}}){Banzatti}, {Pontoppidan},
  {P{\'e}re Ch{\'a}vez}, {Salyk}, {Diehl}, {Bruderer}, {Herczeg}, {Carmona},
  {Pascucci}, {Brittain}, {Jensen}, {Grant}, {van Dishoeck}, {Kamp}, {Bosman},
  {{\"O}berg}, {Blake}, {Meyer}, {Gaidos}, {Boogert}, {Rayner}, \&
  {Wheeler}}]{banz23}
{Banzatti}, A., {Pontoppidan}, K.~M., {P{\'e}re Ch{\'a}vez}, J., {et~al.}
  2023{\natexlab{a}}, \aj, 165, 72, \dodoi{10.3847/1538-3881/aca80b}

\bibitem[{{Banzatti} {et~al.}(2023{\natexlab{b}}){Banzatti}, {Pontoppidan},
  {Carr}, {Jellison}, {Pascucci}, {Najita}, {Mu{\~n}oz-Romero}, {{\"O}berg},
  {Kalyaan}, {Pinilla}, {Krijt}, {Long}, {Lambrechts}, {Rosotti}, {Herczeg},
  {Salyk}, {Zhang}, {Bergin}, {Ballering}, {Meyer}, {Bruderer}, \& {Jdiscs
  Collaboration}}]{banz23b}
{Banzatti}, A., {Pontoppidan}, K.~M., {Carr}, J.~S., {et~al.}
  2023{\natexlab{b}}, \apjl, 957, L22, \dodoi{10.3847/2041-8213/acf5ec}

\bibitem[{{Bast} {et~al.}(2011){Bast}, {Brown}, {Herczeg}, {van Dishoeck}, \&
  {Pontoppidan}}]{bast11}
{Bast}, J.~E., {Brown}, J.~M., {Herczeg}, G.~J., {van Dishoeck}, E.~F., \&
  {Pontoppidan}, K.~M. 2011, \aap, 527, A119,
  \dodoi{10.1051/0004-6361/201015225}

\bibitem[{Giordano {et~al.}(2018)Giordano, Racero, Norman, Vallés, Merín,
  Baines, López-Caniego, Martí, de~Teodoro, Salgado, Sarmiento,
  Gutiérrez-Sánchez, Prieto, Lorca, Alberola, Valtchanov, de~Marchi,
  Álvarez, \& Arviset}]{Giordano_2018}
Giordano, F., Racero, E., Norman, H., {et~al.} 2018, Astronomy and Computing,
  24, 97–103, \dodoi{10.1016/j.ascom.2018.05.002}

\bibitem[{{Goldsmith} \& {Langer}(1999)}]{popdiagram}
{Goldsmith}, P.~F., \& {Langer}, W.~D. 1999, \apj, 517, 209,
  \dodoi{10.1086/307195}

\bibitem[{{Gordon} {et~al.}(2022){Gordon}, {Rothman}, {Hargreaves}, {Hashemi},
  {Karlovets}, {Skinner}, {Conway}, {Hill}, {Kochanov}, {Tan}, {Wcis{\l}o},
  {Finenko}, {Nelson}, {Bernath}, {Birk}, {Boudon}, {Campargue}, {Chance},
  {Coustenis}, {Drouin}, {Flaud}, {Gamache}, {Hodges}, {Jacquemart}, {Mlawer},
  {Nikitin}, {Perevalov}, {Rotger}, {Tennyson}, {Toon}, {Tran}, {Tyuterev},
  {Adkins}, {Baker}, {Barbe}, {Can{\`e}}, {Cs{\'a}sz{\'a}r}, {Dudaryonok},
  {Egorov}, {Fleisher}, {Fleurbaey}, {Foltynowicz}, {Furtenbacher}, {Harrison},
  {Hartmann}, {Horneman}, {Huang}, {Karman}, {Karns}, {Kassi}, {Kleiner},
  {Kofman}, {Kwabia-Tchana}, {Lavrentieva}, {Lee}, {Long}, {Lukashevskaya},
  {Lyulin}, {Makhnev}, {Matt}, {Massie}, {Melosso}, {Mikhailenko}, {Mondelain},
  {M{\"u}ller}, {Naumenko}, {Perrin}, {Polyansky}, {Raddaoui}, {Raston},
  {Reed}, {Rey}, {Richard}, {T{\'o}bi{\'a}s}, {Sadiek}, {Schwenke},
  {Starikova}, {Sung}, {Tamassia}, {Tashkun}, {Vander Auwera}, {Vasilenko},
  {Vigasin}, {Villanueva}, {Vispoel}, {Wagner}, {Yachmenev}, \&
  {Yurchenko}}]{hitran20}
{Gordon}, I.~E., {Rothman}, L.~S., {Hargreaves}, R.~J., {et~al.} 2022, \jqsrt,
  277, 107949, \dodoi{10.1016/j.jqsrt.2021.107949}

\bibitem[{{Grant} {et~al.}(2023){Grant}, {van Dishoeck}, {Tabone}, {Gasman},
  {Henning}, {Kamp}, {G{\"u}del}, {Lagage}, {Bettoni}, {Perotti},
  {Christiaens}, {Samland}, {Arabhavi}, {Argyriou}, {Abergel}, {Absil},
  {Barrado}, {Boccaletti}, {Bouwman}, {o Garatti}, {Geers}, {Glauser},
  {Guadarrama}, {Jang}, {Kanwar}, {Lahuis}, {Morales-Calder{\'o}n}, {Mueller},
  {Nehm{\'e}}, {Olofsson}, {Pantin}, {Pawellek}, {Ray}, {Rodgers-Lee},
  {Scheithauer}, {Schreiber}, {Schwarz}, {Temmink}, {Vandenbussche},
  {Vlasblom}, {Waters}, {Wright}, {Colina}, {Greve}, {Justannont}, \&
  {{\"O}stlin}}]{grant23}
{Grant}, S.~L., {van Dishoeck}, E.~F., {Tabone}, B., {et~al.} 2023, \apjl, 947,
  L6, \dodoi{10.3847/2041-8213/acc44b}

\bibitem[{{Hinkel} {et~al.}(2014){Hinkel}, {Timmes}, {Young}, {Pagano}, \&
  {Turnbull}}]{Hinkel14}
{Hinkel}, N.~R., {Timmes}, F.~X., {Young}, P.~A., {Pagano}, M.~D., \&
  {Turnbull}, M.~C. 2014, \aj, 148, 54, \dodoi{10.1088/0004-6256/148/3/54}

\bibitem[{{Houck} {et~al.}(2004){Houck}, {Roellig}, {van Cleve}, {Forrest},
  {Herter}, {Lawrence}, {Matthews}, {Reitsema}, {Soifer}, {Watson}, {Weedman},
  {Huisjen}, {Troeltzsch}, {Barry}, {Bernard-Salas}, {Blacken}, {Brandl},
  {Charmandaris}, {Devost}, {Gull}, {Hall}, {Henderson}, {Higdon}, {Pirger},
  {Schoenwald}, {Sloan}, {Uchida}, {Appleton}, {Armus}, {Burgdorf},
  {Fajardo-Acosta}, {Grillmair}, {Ingalls}, {Morris}, \& {Teplitz}}]{irs}
{Houck}, J.~R., {Roellig}, T.~L., {van Cleve}, J., {et~al.} 2004, \apjs, 154,
  18, \dodoi{10.1086/423134}

\bibitem[{{Jellison} {et~al.}(2024){Jellison}, {Johnson}, {Banzatti}, \&
  {Bruderer}}]{iSLAT}
{Jellison}, E., {Johnson}, M., {Banzatti}, A., \& {Bruderer}, S. 2024, arXiv
  e-prints, arXiv:2402.04060, \dodoi{10.48550/arXiv.2402.04060}

\bibitem[{{Kaeufl} {et~al.}(2004){Kaeufl}, {Ballester}, {Biereichel},
  {Delabre}, {Donaldson}, {Dorn}, {Fedrigo}, {Finger}, {Fischer}, {Franza},
  {Gojak}, {Huster}, {Jung}, {Lizon}, {Mehrgan}, {Meyer}, {Moorwood}, {Pirard},
  {Paufique}, {Pozna}, {Siebenmorgen}, {Silber}, {Stegmeier}, \&
  {Wegerer}}]{crires}
{Kaeufl}, H.-U., {Ballester}, P., {Biereichel}, P., {et~al.} 2004, in Society
  of Photo-Optical Instrumentation Engineers (SPIE) Conference Series, Vol.
  5492, Ground-based Instrumentation for Astronomy, ed. A.~F.~M. {Moorwood} \&
  M.~{Iye}, 1218--1227, \dodoi{10.1117/12.551480}

\bibitem[{{Lagage} {et~al.}(2004){Lagage}, {Pel}, {Authier}, {Belorgey},
  {Claret}, {Doucet}, {Dubreuil}, {Durand}, {Elswijk}, {Girardot}, {K{\"a}ufl},
  {Kroes}, {Lortholary}, {Lussignol}, {Marchesi}, {Pantin}, {Peletier},
  {Pirard}, {Pragt}, {Rio}, {Schoenmaker}, {Siebenmorgen}, {Silber}, {Smette},
  {Sterzik}, \& {Veyssiere}}]{visir}
{Lagage}, P.~O., {Pel}, J.~W., {Authier}, M., {et~al.} 2004, The Messenger,
  117, 12

\bibitem[{{Mace} {et~al.}(2016){Mace}, {Kim}, {Jaffe}, {Park}, {Lee}, {Kaplan},
  {Yu}, {Yuk}, {Chun}, {Pak}, {Kim}, {Lee}, {Sneden}, {Afsar}, {Pavel}, {Lee},
  {Oh}, {Jeong}, {Park}, {Kidder}, {Lee}, {Nguyen Le}, {McLane},
  {Gully-Santiago}, {Oh}, {Lee}, {Hwang}, \& {Park}}]{igrins}
{Mace}, G., {Kim}, H., {Jaffe}, D.~T., {et~al.} 2016, in Society of
  Photo-Optical Instrumentation Engineers (SPIE) Conference Series, Vol. 9908,
  Ground-based and Airborne Instrumentation for Astronomy VI, ed. C.~J.
  {Evans}, L.~{Simard}, \& H.~{Takami}, 99080C, \dodoi{10.1117/12.2232780}

\bibitem[{{Mason} {et~al.}(2001){Mason}, {Wycoff}, {Hartkopf}, {Douglass}, \&
  {Worley}}]{WDS01}
{Mason}, B.~D., {Wycoff}, G.~L., {Hartkopf}, W.~I., {Douglass}, G.~G., \&
  {Worley}, C.~E. 2001, \aj, 122, 3466, \dodoi{10.1086/323920}

\bibitem[{{McLean} {et~al.}(1998){McLean}, {Becklin}, {Bendiksen}, {Brims},
  {Canfield}, {Figer}, {Graham}, {Hare}, {Lacayanga}, {Larkin}, {Larson},
  {Levenson}, {Magnone}, {Teplitz}, \& {Wong}}]{nirspec}
{McLean}, I.~S., {Becklin}, E.~E., {Bendiksen}, O., {et~al.} 1998, in Society
  of Photo-Optical Instrumentation Engineers (SPIE) Conference Series, Vol.
  3354, Infrared Astronomical Instrumentation, ed. A.~M. {Fowler}, 566--578,
  \dodoi{10.1117/12.317283}

\bibitem[{{Pontoppidan} {et~al.}(2024){Pontoppidan}, {Salyk}, {Banzatti},
  {Zhang}, {Pascucci}, {{\"O}berg}, {Long}, {Mu{\~n}oz-Romero}, {Carr},
  {Najita}, {Blake}, {Arulanantham}, {Andrews}, {Ballering}, {Bergin},
  {Calahan}, {Cobb}, {Colmenares}, {Dickson-Vandervelde}, {Dignan}, {Green},
  {Heretz}, {Herczeg}, {Kalyaan}, {Krijt}, {Pauly}, {Pinilla}, {Trapman}, \&
  {Xie}}]{pontoppidan24}
{Pontoppidan}, K.~M., {Salyk}, C., {Banzatti}, A., {et~al.} 2024, \apj, 963,
  158, \dodoi{10.3847/1538-4357/ad20f0}

\bibitem[{{Ram{\'\i}rez-Tannus} {et~al.}(2023){Ram{\'\i}rez-Tannus}, {Bik},
  {Cuijpers}, {Waters}, {G{\"o}ppl}, {Henning}, {Kamp}, {Preibisch}, {Getman},
  {Chaparro}, {Cuartas-Restrepo}, {de Koter}, {Feigelson}, {Grant}, {Haworth},
  {Hern{\'a}ndez}, {Kuhn}, {Perotti}, {Povich}, {Reiter}, {Roccatagliata},
  {Sabbi}, {Tabone}, {Winter}, {McLeod}, {van Boekel}, \& {van
  Terwisga}}]{ramireztannus23}
{Ram{\'\i}rez-Tannus}, M.~C., {Bik}, A., {Cuijpers}, L., {et~al.} 2023, \apjl,
  958, L30, \dodoi{10.3847/2041-8213/ad03f8}

\bibitem[{{Rayner} {et~al.}(2022){Rayner}, {Tokunaga}, {Jaffe}, {Bond},
  {Bonnet}, {Ching}, {Connelley}, {Cushing}, {Kokubun}, {Lockhart}, {Vacca}, \&
  {Warmbier}}]{ishell}
{Rayner}, J., {Tokunaga}, A., {Jaffe}, D., {et~al.} 2022, \pasp, 134, 015002,
  \dodoi{10.1088/1538-3873/ac3cb4}

\bibitem[{{Rieke} {et~al.}(2015){Rieke}, {Wright}, {B{\"o}ker}, {Bouwman},
  {Colina}, {Glasse}, {Gordon}, {Greene}, {G{\"u}del}, {Henning}, {Justtanont},
  {Lagage}, {Meixner}, {N{\o}rgaard-Nielsen}, {Ray}, {Ressler}, {van Dishoeck},
  \& {Waelkens}}]{miri}
{Rieke}, G.~H., {Wright}, G.~S., {B{\"o}ker}, T., {et~al.} 2015, \pasp, 127,
  584, \dodoi{10.1086/682252}

\bibitem[{Stassun {et~al.}(2019)Stassun, Oelkers, Paegert, Torres, Pepper, Lee,
  Collins, Latham, Muirhead, Chittidi, Rojas-Ayala, Fleming, Rose, Tenenbaum,
  Ting, Kane, Barclay, Bean, Brassuer, Charbonneau, Ge, Lissauer, Mann, McLean,
  Mullally, Narita, Plavchan, Ricker, Sasselov, Seager, Sharma, Shiao,
  Sozzetti, Stello, Vanderspek, Wallace, \& Winn}]{StassunOelkers_2019}
Stassun, K.~G., Oelkers, R.~J., Paegert, M., {et~al.} 2019, The Astronomical
  Journal, 158, 138, \dodoi{10.3847/1538-3881/ab3467}

\bibitem[{{Temmink} {et~al.}(2024){Temmink}, {van Dishoeck}, {Grant}, {Tabone},
  {Gasman}, {Christiaens}, {Samland}, {Argyriou}, {Perotti}, {G{\"u}del},
  {Henning}, {Lagage}, {Abergel}, {Absil}, {Barrado}, {Caratti o Garatti},
  {Glauser}, {Kamp}, {Lahuis}, {Olofsson}, {Ray}, {Scheithauer},
  {Vandenbussche}, {Waters}, {Arabhavi}, {Jang}, {Kanwar},
  {Morales-Calder{\'o}n}, {Rodgers-Lee}, {Schreiber}, {Schwarz}, \&
  {Colina}}]{Temmink24}
{Temmink}, M., {van Dishoeck}, E.~F., {Grant}, S.~L., {et~al.} 2024, \aap, 686,
  A117, \dodoi{10.1051/0004-6361/202348911}

\bibitem[{{Wells} {et~al.}(2015){Wells}, {Pel}, {Glasse}, {Wright},
  {Aitink-Kroes}, {Azzollini}, {Beard}, {Brandl}, {Gallie}, {Geers}, {Glauser},
  {Hastings}, {Henning}, {Jager}, {Justtanont}, {Kruizinga}, {Lahuis}, {Lee},
  {Martinez-Delgado}, {Mart{\'\i}nez-Galarza}, {Meijers}, {Morrison},
  {M{\"u}ller}, {Nakos}, {O'Sullivan}, {Oudenhuysen}, {Parr-Burman}, {Pauwels},
  {Rohloff}, {Schmalzl}, {Sykes}, {Thelen}, {van Dishoeck}, {Vandenbussche},
  {Venema}, {Visser}, {Waters}, \& {Wright}}]{jwst-mrs}
{Wells}, M., {Pel}, J.~W., {Glasse}, A., {et~al.} 2015, \pasp, 127, 646,
  \dodoi{10.1086/682281}

\bibitem[{{Wenger} {et~al.}(2000){Wenger}, {Ochsenbein}, {Egret}, {Dubois},
  {Bonnarel}, {Borde}, {Genova}, {Jasniewicz}, {Lalo{\"e}}, {Lesteven}, \&
  {Monier}}]{Wenger2000}
{Wenger}, M., {Ochsenbein}, F., {Egret}, D., {et~al.} 2000, \aaps, 143, 9,
  \dodoi{10.1051/aas:2000332}

\bibitem[{{Wright} {et~al.}(2023){Wright}, {Rieke}, {Glasse}, {Ressler},
  {Garc{\'\i}a Mar{\'\i}n}, {Aguilar}, {Alberts}, {{\'A}lvarez-M{\'a}rquez},
  {Argyriou}, {Banks}, {Baudoz}, {Boccaletti}, {Bouchet}, {Bouwman}, {Brandl},
  {Breda}, {Bright}, {Cale}, {Colina}, {Cossou}, {Coulais}, {Cracraft}, {De
  Meester}, {Dicken}, {Engesser}, {Etxaluze}, {Fox}, {Friedman}, {Fu},
  {Gasman}, {G{\'a}sp{\'a}r}, {Gastaud}, {Geers}, {Glauser}, {Gordon},
  {Greene}, {Greve}, {Grundy}, {G{\"u}del}, {Guillard}, {Haderlein},
  {Hashimoto}, {Henning}, {Hines}, {Holler}, {Detre}, {Jahromi}, {James},
  {Jones}, {Justtanont}, {Kavanagh}, {Kendrew}, {Klaassen}, {Krause},
  {Labiano}, {Lagage}, {Lambros}, {Larson}, {Law}, {Lee}, {Libralato}, {Lorenzo
  Alverez}, {Meixner}, {Morrison}, {Mueller}, {Murray}, {Mycroft}, {Myers},
  {Nayak}, {Naylor}, {Nickson}, {Noriega-Crespo}, {{\"O}stlin}, {O'Sullivan},
  {Ottens}, {Patapis}, {Penanen}, {Pietraszkiewicz}, {Ray}, {Regan},
  {Roteliuk}, {Royer}, {Samara-Ratna}, {Samuelson}, {Sargent}, {Scheithauer},
  {Schneider}, {Schreiber}, {Shaughnessy}, {Sheehan}, {Shivaei}, {Sloan},
  {Tamas}, {Teague}, {Temim}, {Tikkanen}, {Tustain}, {van Dishoeck},
  {Vandenbussche}, {Weilert}, {Whitehouse}, \& {Wolff}}]{miri2}
{Wright}, G.~S., {Rieke}, G.~H., {Glasse}, A., {et~al.} 2023, \pasp, 135,
  048003, \dodoi{10.1088/1538-3873/acbe66}

\end{thebibliography}
\end{document}